\documentclass[12pt,epsfig]{article}
\usepackage{amsfonts,latexsym,graphicx}
\DeclareMathAlphabet{\mathpzc}{OT1}{pzc}{m}{it}  
\begin{document}

\title{Algebraic Bethe ansatz method for the exact calculation of energy 
spectra and form factors: applications to 
models of Bose-Einstein condensates and
metallic nanograins}

\author{Jon Links\footnote{email: jrl@maths.uq.edu.au}, 
Huan-Qiang Zhou\footnote{email: hqz@maths.uq.edu.au}, 
Ross H. McKenzie and Mark D. Gould  
\\ Centre for Mathematical Physics, School of Physical Sciences,
\\
The University of Queensland, Brisbane 4072, Australia.}
\maketitle
\begin{abstract}
In this review we demonstrate how the algebraic Bethe ansatz 
is used for the calculation of the energy spectra and 
form factors (operator matrix elements in the basis of Hamiltonian 
eigenstates) in exactly solvable quantum systems. As examples we apply
the theory to several models of current interest in the 
study of Bose-Einstein condensates, which have been 
successfully created using ultracold dilute atomic gases. The first model we
introduce describes
Josephson tunneling between two coupled  Bose-Einstein
condensates. It can be used not only for the study of tunneling between
condensates of atomic gases, but for solid state Josephson junctions
and coupled Cooper pair boxes. The theory is also applicable to models 
of atomic-molecular Bose-Einstein 
condensates, with two examples given and analysed. 
Additionally, these same two models are relevant to studies in quantum optics. 
Finally, we discuss the 
model of Bardeen, Cooper and Schrieffer 
in this framework, which is appropriate for systems of ultracold 
fermionic atomic gases, as well as being applicable  
for the description of superconducting correlations in 
metallic grains with nanoscale dimensions. 
In applying all of the above models to physical situations, 
the need for an exact analysis of small scale systems is 
established due to large quantum fluctuations which render mean-field
approaches inaccurate. 
\end{abstract}



\vfil\eject
\tableofcontents 
\vfil\eject



\def\b{\beta}
\def\d{\dagger}
\def\e{\epsilon}
\def\K{\kappa}
\def\l{\left}
\def\r{\right} 
\def\o{\omega}
\def\t{\theta}
\def\s{\sigma}
\def\ap{\approx} 


\def\beq{\begin{equation}}
\def\eeq{\end{equation}}
\def\bea{\begin{eqnarray}}
\def\eea{\end{eqnarray}}
\def\ba{\begin{eqnarray}}
\def\ea{\end{eqnarray}}
\def\no{\nonumber}
\def\le{\langle}
\def\re{\rangle}
\def\lt{\left}
\def\rt{\right}
\def\o{\omega}
\def\d{\dagger}
\def\nn{\nonumber} 
\def\j{{ {\cal J}}}
\def\n{{\hat n}}
\def\N{{\hat N}}
\def\A{\mathcal{A}}
\def\B{\mathcal{B}}
\def\C{\mathcal{C}}
\def\D{\mathcal{D}}
\def\R{\mathcal{R}}
\def\Y{\mathcal{Y}}
\def\Z{{\mathbb{Z}}}
\def\I{\mathcal{I}}
\def\T{\mathcal{T}}
\def\E{\mathcal{E}}
\def\G{\mathcal{G}}
\def\GG{\overline{\mathcal{G}}}
\def\FF{\overline{\mathcal{F}}}
\def\FFF{\overline{\overline{\mathcal{F}}}}
\def\P{\mathcal{P}}
\def\PP{\overline{\mathcal{P}}}
\def\S{\mathcal{S}}
\def\L{\mathcal{L}}
\def\W{\mathcal{W}}
\def\F{\mathcal{F}}
\def\KK{\mathcal{K}}
\def\KKK{\overline{\mathcal{K}}}
\def\Q{\mathcal{Q}}
\def\QQ{\overline{\mathcal{Q}}}
\def\aa{{\mathpzc{a}}}
\def\dd{{\mathpzc{d}}}
\def\g{{\mathfrak{g}}}
\def\cC{{\mathfrak{C}}}
\def\Y{\mathcal{Y}}
\def\j{\mathcal{J}} 
\def\n{{\hat n}}
\def\N{{\hat N}}
\def\La{\Lambda}
\def\rmd{{\rm{d}}} 

\section{Introduction}

Exactly solvable models of quantum mechanical systems provide 
an important 
insight into the nature of quantum physics, with the simple harmonic
oscillator and the non-relativistic hydrogen atom serving as the
archetypal examples. One method to solve these models is to exploit an
underlying algebraic structure, well known to be 
the Lie algebra $gl(3)$ for the
harmonic oscillator and $so(4)$ for the hydrogen model \cite{wybourne}. 
In this approach, 
the Lie algebraic structure plays the role of generating states of the
system while at the same time providing state labels (or
quantum numbers).
A celebrated exact solution of a quantum many-body model
is that for the 
one-dimensional Heisenberg (spin 1/2) chain, due to Bethe \cite{bethe}. 
Out of this work grew the concept of the Bethe ansatz for the
construction of the eigenvectors for an exactly solvable Hamiltonian.  
In adopting this method, a general possible form for an  
eigenvector is assumed, dependent on several free parameters. Constraints
are then determined for the parameters which ensure that this vector is
an eigenvector of the Hamiltonian. 
The constraint equations are referred to as the Bethe
ansatz equations of the model. 

Motivated by Bethe's work the field of exactly solvable models
flourished during the 1960s led by McGuire \cite{mcguire}, 
Lieb \cite{lieb}, Sutherland \cite{sutherland}, 
Yang \cite{yang} and Baxter \cite{baxter}, 
amongst many others. Out of this activity arose the
Yang-Baxter equation, the solution of which provides a sufficiency  
condition to construct a model which is exactly solvable (applicable to
one-dimensional quantum spin chains, including quantum field theories as the
lattice spacing goes to zero, and classical two-dimensional  
lattice systems) \cite{knot1,int}.   
A fundamental feature of the Yang-Baxter equation is that 
it can always be used to construct a family of mutually commuting
matrices, known as transfer matrices, 
which facilitates the application of the Bethe
ansatz. The method of the Bethe ansatz can take a variety of forms,
commonly known as the co-ordinate, analytic, functional
and algebraic forms. It is this latter approach that will be the
focus of our work here, as this is the most appropriate to serve our
requirements. 

The algebraic formulation of the Bethe ansatz, and the associated
quantum inverse scattering method, was 
primarily developed by the group of
mathematical physicists in St. Petersburg 
\cite{fst,ks,takhtajan,korepin,faddeev}. 
Its applicability extends beyond the study of 
one-dimensional spin chains, quantum field theory 
and two-dimensional lattice models to 
systems of correlated electrons \cite{ek}, conformal field theory
\cite{blz}, as well as
precipitating the notion of quantum algebras (deformations of universal
enveloping algebras of Lie algebras)
\cite{jimbo85,jimbo86,drinfeld,frt}.  
The main motivation for the algebraic formulation of the Bethe
ansatz was not only for calculating the energy spectrum of a model, but to also 
accommodate the calculation of correlation functions. An
initial step in this direction is to compute the form factors of an
operator (not necessarily observable). Here the term ``form factors of
an operator'' simply
refers to the matrix elements of that operator in the basis of
Hamiltonian 
eigenstates. Expectation values of observable operators and
general correlation functions are expressible in terms of form
factors through completeness relations. 

The study of correlation
functions in the context of exactly solvable models 
has its origins in Baxter's corner transfer matrix method
\cite{baxter}. Following on from this there has been a rich theory developed 
using ideas taken from affine quantum algebras, vertex operators, integrable
field theories, the 
off-shell Bethe ansatz and the Knizhnik-Zamolodchikov equation 
\cite{fr,smirnov,dfjmn,babujian,jm,karowski1,karowski2}, 
as well as the algebraic Bethe ansatz
approach (e.g \cite{korepin,efik,kmt,ks99,sklyanin2}).    
For the models we will study here  
the calculation of form factors will be undertaken through extensive use
of the Slavnov formula \cite{slavnov} for the scalar products of Bethe
eigenstates. The Slavnov formula provides an explicit determinant representation for
the scalar products. A refined proof of this result was given
by Kitanine, Maillet and Terras \cite{kmt}, using the notion of
factorising the solutions of the Yang-Baxter equation 
in terms of Drinfeld twists \cite{twist}. 
They applied
this method to successfully compute form factors for the
anisotropic ($XXZ$) Heisenberg chain \cite{kmt}, and in a closely related
work Korepin and Slavnov computed form factors for the quantum
non-linear Schr\"odinger equation \cite{ks99}. In both cases the results
are valid for finite size systems, and thus this approach is appropriate
for applications to nanoscale systems. The results presented here
are largely inspired by these works.  

The need to appeal to the exact solution of a model has been
well illustrated in the context of the energy spectrum of metallic grains of
nanoscale size. Experiments conducted by Ralph, Black and Tinkham (RBT)
\cite{brt,rbt} using single electron tunneling spectroscopy on aluminium grains
with mean radii in the range 5-13 nm 
indicated significant parity effects due to the
number of electrons in the system.  The electron number 
remains fixed due to the large
charging energy of the grains, which is a consequence of
their small size.  For grains with an odd number of electrons,
the gap in the energy spectrum reduces with increasing size of the system, in
contrast
to the case of a grain with an even number of electrons, where a gap
larger than the single electron energy levels persists. In the latter
case the gap can be closed by a strongly applied magnetic field.
The conclusion drawn from these results is that pairing interactions
are prominent in these nanoscale systems.
For a grain with an odd number of electrons there will always
be at least one unpaired electron, so it is not necessary to break a
Cooper pair in order to create an excited state. For a grain with an even
number
of electrons, all excited states have a least one broken Cooper pair,
resulting in a gap in the spectrum.
In the presence of a
strongly applied magnetic field, it is energetically more favourable
for a grain with an even number of electrons
to have broken pairs, and hence in this case
there are excitations which show no gap in the
spectrum.

A na\"{\i}ve
approach to describe
these nanograins is to apply the theory of
superconductivity
due to Bardeen, Cooper and Schrieffer (BCS) \cite{bcs}.
Indeed, the BCS model is appropriate for these systems
but the associated
mean-field treatment fails. There are two main reasons for this. First
is because the BCS analysis makes use of the grand canonical ensemble
whereas in the experiments the electron number is fixed. 
Second is because a mean-field theory
approximates
certain operators in the model by an average value. At the nanoscale
level, the quantum fluctuations are sufficiently large enough that this
approximation is invalid.
In systems where the
mean single particle energy level spacing, which is inversely
proportional to the volume, is comparable to the bulk superconducting gap,
(as is for metallic nanograins,) it was thought that pairing interactions
would not correlate any energy levels.
This was conjectured by Anderson \cite{anderson} 
on the basis of the BCS mean-field analysis,
but the experiments of RBT show this to not be the case.
Consequently, an exact solution was desired in order to clarify the
issue. 

Remarkably, the exact solution of the reduced BCS model (``reduced'' refers
to the fact that only zero momentum Cooper pairs are considered and 
all couplings for scattering of Cooper pairs are equal) 
had been
obtained and analysed many years earlier in a series of works by
Richardson and Sherman \cite{richardson,rs} using an approach equivalent
to the co-ordinate Bethe ansatz.  
The motivation for their work was for application of pairing
interactions in nuclear
systems, which involve fixed particle number, and thus 
the BCS analysis referred above is not valid.   
However, the condensed matter physics
community was unaware of this earlier work at the time  
the results of RBT were communicated. 
It was subsequently shown that theoretical results obtained through 
an analysis of the exact solution 
for the reduced 
BCS Hamiltonian were compatible with the experimental results of RBT
\cite{vdr}.

One of the most currently active fields is the study of Bose-Einstein
condensates of ultracold atomic gases 
\cite{cw,ak}. The
Bose-Einstein condensed state is of a purely quantum mechanical nature and,  
in analogy with the phenomena of superconducting metallic nanograins
discussed above, a mean-field analysis of small scale systems comprised
of  Bose-Einstein condensates is inadequate due to
significant quantum fluctuations. 
While there are many novel physical properties to be explored in the
study of Bose-Einstein condensates, there are three features that we will 
discuss here. The first is the phenomenon of Josephson tunneling between
two coupled Bose-Einstein condensates. Recall that the Josephson effect
was first proposed in relation to the tunneling of Cooper pairs through
an insulating barrier separating two superconductors 
\cite{josephson1,josephson2}. (A very informative historical account is
given in \cite{mcdonald}.)
It has been proposed as a means to couple qubits for the purpose
of quantum computation \cite{mss,leggett2}. 
The experimental realisation of Bose-Einstein condensation in the atomic
alkali gases provides a framework in which to observe macroscopic
tunneling in a system with tunable couplings. 
An extensive account of this
phenomenon can be found in \cite{leggett}, which discusses in detail the
canonical Josephson Hamiltonian (equivalent to a two site Bose-Hubbard
model) for the description of this effect. 
It is not well known that this model is exactly solvable
through the quantum inverse scattering method,
which was established about a decade ago in the context of the
discrete self-trapping dimer model \cite{enolskii1,enolskii2}. 
Below we show that a
slightly more general model is also exactly solvable and we derive
explicit exact form factors for the generalised model.

The second aspect of Bose-Einstein condensation we will discuss is
that of a condensate comprised of a coherent superposition of atomic and
molecular states. 
This phenomenon has been predicted and studied by theorists (e.g., see 
\cite{dkh,java,hwdk,vya,caok}) 
and recently realised experimentally \cite{wfhrh,z,dctw}. 
In particular, for the experiment of \cite{dctw} using $^{85}$Rb
atoms, which are converted into 
diatomic molecules via a Feshbach resonance, the system was
prepared, allowed to evolve, and then a measurement made to 
determine the number of
atoms in the system. By performing this procedure over different
evolution times it was established that the expectation value for the
number of atoms displayed an oscillatory behaviour, indicating that
the state of the system was a quantum mechanical superposition 
of atomic and molecular states, as opposed to a classical mixture. 
The result is significant in that the state of the system is comprised of
a superposition of two chemically distinct components. 

Finally, we will also analyse the reduced BCS model, which is relevant not
only for metallic nanograins as described above, but also the
study of ultracold fermionic atomic gases \cite{jin}. 
As is well known, for an ultracold fermionic gas the Pauli principle
prohibits all particles occupying the lowest energy level. The
lowest possible energy of the system is obtained by filling the Fermi sea.   
However, in analogy with metals it is believed that fermionic gases
should be able to form Cooper pairs, and as a consequence, undergo a
phase transition at a suitably low temperature into a fermionic
condensate \cite{tsmph}. 

The aim of this exposition is to illustrate 
that the algebraic approach to the study of exactly solvable models is 
a rich and elegant theory with wide applicability. 
In particular we show how the theory applies to the systems of
Bose-Einstein condensates and the reduced BCS model discussed above. 
Some of these results have already been communicated 
\cite{zlmg,lz,flz,zlmx,zlm,zlgm}, while other results we will present
are new. 
In each case we will determine the energy spectrum, as well as the form
factors for the computation of correlation functions,
in terms of the Bethe ansatz solution. 
Certain correlation functions can in fact be deduced
directly from the energy spectrum through use of the Hellmann-Feynman
theorem \cite{h,f}. Examples of this procedure applied to the models
discussed here can be found in \cite{flz,zlmx,zlm}. Typically however, the
form factor approach is required to build general formulae for
expectation values and correlation
functions. As a potential
application of these results we point to the problem of quantifying
entanglement in the theory of quantum information. The role of
correlation functions in the characterisation of entanglement has been
discussed in \cite{on,oaff,kter,hmm}. 

Throughout, we have endeavoured to provide
as much technical detail as possible for the benefit of non-experts. 
The exceptions are  the Slavnov
formula for the scalar product of states, the proof of which is beyond
the scope of this review. For the proof we refer the interested reader
to \cite{kmt}.
Also, the orthogonality of the Bethe eigenstates will not be
proved. Details of this result can be found in \cite{ks99}. 
The format of the review is as follows. 
We begin in section 2 with a description of the four models we will
examine. 
In section 3 we recall the basic features of the quantum inverse scattering
method for the construction of exactly solvable models. 
While there already exist several excellent surveys of this approach 
\cite{fst,ks,takhtajan,korepin,faddeev}, 
we give a detailed account here in order to
fix notations and conventions and make the review self-contained. 
The central aspect is the introduction
of the Yang-Baxter algebra associated with the Lie algebra $gl(2)$, 
which is a quadratic algebra. Several examples of realisations are 
given.  We show that in a particular
limit, called the quasi-classical limit, 
the Yang-Baxter algebra reduces to a Lie algebra, called
the Gaudin algebra. Through a realisation of the
Yang-Baxter algebra, the transfer matrix is constructed which leads to
an exactly solvable model. 
We also discuss a natural $\Z$-graded structure of the Yang-Baxter algebra
which will be exploited in later constructions.     
Section 4 deals with the algebraic Bethe ansatz method in a general
context for the
determination of the spectrum of the transfer matrix. 
Section 5  presents the Slavnov formula for the scalar products of
the states which arise in the algebraic Bethe ansatz method of solution.
We also discuss how, through the use of the Slavnov formula, the form
factors for the elements of the Yang-Baxter algebra can be obtained.  
Section 6 turns to calculating 
the explicit exact solutions for the models. 
Formulae for the energy spectrum are determined, 
which are parameterised in terms of the roots of the Bethe ansatz
equations. 
Section 7 deals with the computation of form factors for each of the
models introduced. In all cases it is necessary to first consider
the solution to the inverse problem, which involves expressing a given
operator in terms of the elements of the Yang-Baxter algebra. This needs
to be studied on a case by case basis. Once this is achieved, the form
factors for that operator can be determined. Concluding remarks are given
in section 8. 

\section{Model Hamiltonians} 

Here we present, and give a description of, three models for Bose-Einstein
condensates and the reduced BCS model. Our main objective is to
establish that each model is exactly solvable through the algebraic
Bethe ansatz. Throughout, there are no constraints imposed on the
coupling parameters for all models other than they are real, which is to
ensure hermiticity. 

\subsection{A model for two Josephson coupled Bose-Einstein condensates}

Consider the following general Hamiltonian describing Josephson
tunneling
between two coupled Bose-Einstein condensates
\bea
H&=& U_{11} N_1^2+U_{12}N_1N_2+U_{22} N_2^2 +\mu_1 N_1+\mu_2 N_2\no\\
&& -\frac{\E_J}{2} (a_1^\dagger a_2 + a_2^\dagger a_1)
\label{jo} \eea
where the operators $a_i,\,a^\dagger_i,\,N_i=a^\dagger_i a_i$ are associated
with two Heisenberg algebras with relations 
$$[a_i,\,a_j^\dagger]=\delta_{ij},~~~[a_i,\,a_j]=[a^\dagger_i,\,a^\dagger_j]
=0. $$   
The Hilbert space of states is given by the infinite-dimensional Fock space
spanned by the vectors
\beq \left|m,n\right>=(a_1^{\d})^m(a_2^{\d})^n\left|0\right>, ~~~~~~~~
m,\,n=0,1,2,....,\infty . \label{fock} \eeq
The model describes Josephson tunneling
between two condensates with tunneling strength ${\E_J}/{2}$, the
parameters $U_{ij}$ are the amplitudes for $S$-wave scattering and
$\mu_i$ are chemical potentials. The Hamiltonian commutes with the total
particle number $N=N_1+N_2$. 

The above Hamiltonian under the constraint $U=U_{11}=U_{22}=-U_{12}/2$
has been
studied widely
using techniques other than the exact solution
\cite{leggett,mcww,dgps,otfyk,ps,ads}.
For this case it is useful to divide the parameter space into three
regimes; viz. Rabi ($U/\E_J<<N^{-1}$), Josephson
($N^{-1}<<U/\E_J<<N$) and Fock ($N<<U/\E_J$).
In the Rabi and Josephson regions one expects coherent superposition of the two
condensates to be possible whereas in the
Fock
region the two condensates will be, in some sense, localised.
There is a correspondence between (\ref{jo}) and the motion of a
pendulum
\cite{leggett}. In the Rabi and Josephson regions this motion is
semiclassical, (i.e., the energy level spacings are of order less than
$N$,) 
in contrast to the Fock case. For both the Fock and Josephson regimes
the
analogy corresponds to a pendulum with fixed length, while in the Rabi
regime the length varies. An important problem is to study the behaviour
in the
crossover regimes, particularly between the Josephson and Fock regimes
which are the most likely to occur in an experimental context
\cite{leggett}.
A reliable method to do this is through the exact
solution.
The motivation to extend the solution to the case where the couplings
$U_{11},\,
U_{22},\,U_{12}$ for the $S$-wave scattering terms can be chosen
arbitrarily
is for the description of a pair of Cooper pair boxes with
capacitive coupling \cite{mss}. In the limit $U_{22}\rightarrow 0$,
then $\left<N_2\right> >> \left<N_1\right>$,
in which case the model can be considered as a single Cooper
pair box coupled to a reservoir.

\subsection{A model for homo-atomic-molecular Bose-Einstein condensates}

Next we turn our attention to a two-mode model for
an atomic-molecular Bose-Einstein condensate with identical
atoms. The Hamiltonian
takes the form
\bea
H&=& U_{aa} N_a^2+U_{ac}N_aN_c+U_{cc} N_c^2 +\mu_a N_a+\mu_c N_c\no\\
&& +\Omega (a^\dagger a^\dagger c + c^\dagger a a) \label{acham}
\eea
which acts on a basis of Fock states analogous to (\ref{fock}).
Here, $a^\dagger$ is the creation operator for an atomic mode while
$c^\dagger$ creates a molecular mode. The parameters $U_{ij}$ again describe
$S$-wave scattering, $\mu_i$ are chemical potentials and $\Omega$ is the
amplitude for interconversion of atoms and molecules. 
The Hamiltonian commutes with the total atom number $N=N_a+2N_c$.

In the limit $U_{aa}=U_{ac}=U_{cc}=0$ this model was studied 
in \cite{vya}, and analysed  numerically  in \cite{zlm}
based on the Bethe ansatz solution. However, in order to compare with 
experimental results, in which  
the $S$-wave scatterings are significant, one needs to analyse
(\ref{acham}) in its full generality. Estimates for the $S$-wave
scattering parameters in the case of $^{87}$Rb are given in \cite{caok}.

\subsection{A model for hetero-atomic-molecular Bose-Einstein
condensates}

The previous model can be extended to describe an
atomic-molecular Bose-Einstein condensate with two distinct
species of
atoms, denoted $a$ and $b$, which can combine to produce a molecule $c$. 
For this case the Hamiltonian
takes the form
\bea
H&=& U_{aa} N_a^2+U_{bb} N_b^2 +U_{cc} N_c^2+
U_{ab}N_aN_b+U_{ac} N_aN_c+U_{bc}N_bN_c \no\\
&&+\mu_a N_a+\mu_b N_b+\mu_c N_c
+\Omega (a^\dagger b^\dagger c + c^\dagger b a)
\label{abcham} \eea
which commutes with $\I=N_a-N_b$ and the total atom number 
$N=N_a+N_b+2N_c$.  
Here the model acts on the Fock space spanned by the vectors
$$\left|l,m,n\right>=(a^\dagger)^l(b^\dagger)^m(c^\dagger)^n\left|0\right>.
$$
Let us point out that in the limit
$U_{aa}=U_{bb}=U_{cc}=U_{ab}=U_{ac}=U_{bc}=0$, equation  
(\ref{abcham}) is the Hamiltonian studied in \cite{wb,wt}
modelling second harmonic generation in quantum optics.
Non-zero values of these parameters correspond to a Kerr effect.

\subsection{The reduced BCS model}

The physical properties of a metallic nanograin with pairing
interactions are described by the
reduced BCS Hamiltonian \cite{vdr}
\beq H=\sum_{j=1}^{\L}\e_jn_j
-g\sum_{j,k=1}^{\L}c_{k+}^{\d}c_{k-}^{\d}c_{j-}c_{j+}. \label{bcs}
\eeq
Above, $j=1,...,{\L}$ labels a shell of doubly degenerate single
particle
energy levels with energies $\e_j$ and $n_j=c^\dagger_{j+}c_{j+}
+c^\dagger_{j-}c_{j-}$ is the
fermion number operator for
level $j$. The operators $c_{j\pm},\,c^{\d}_{j\pm}$ are the annihilation
and creation operators for the fermions at level $j$. The labels $\pm$
refer
to time reversed states.

One of the features of the Hamiltonian (\ref{bcs})
is the {\it blocking
effect}. For any unpaired fermion at level $j$ the action of
the pairing interaction is zero since only paired fermions are
scattered. This means that the Hilbert space can be decoupled into
a product of paired and unpaired fermion states in which the
action of the Hamiltonian on the space for the unpaired fermions is
automatically diagonal in the natural basis.
In view of the blocking effect, it is convenient to introduce
hard-core boson operators
$b_j=c_{j-}c_{j+},\, b^{\d}_{j}=c^{\d}_{j+}c^{\d}_{j-}$ which satisfy
the relations
\beq 
(b^{\d}_j)^2=0, ~~~~[b_j,\,b_k^{\d}]=\delta_{jk}(1-2b^{\d}_jb_j)
~~~~[b_j,\,b_k]=[b^{\d}_j,\,b^{\d}_k]=0 \label{hcb} \eeq 
on the space excluding single particle states.
In this setting the hard-core boson operators realise the
$su(2)$ algebra in the pseudo-spin representation,
which will be utilised below.

The original approach of Bardeen, Cooper and Schrieffer \cite{bcs}
to describe the
phenomenon of superconductivity in a bulk system was to employ
a mean-field theory using a variational wavefunction for the ground
state
\beq \left|\Psi\right>=\prod_{i=1}^{\L}(u_iI+v_ib_i^{\d})\left|0\right>
\label{bcswf} \eeq
which has an undetermined number of electrons. The expectation value for
the
number operator is then fixed by means of a chemical potential term
$\mu$; i.e. the grand canonical ensemble is used.
One of the
predictions of the BCS theory is that the number of Cooper pairs in the
ground
state of the system is given by the ratio $\Delta/d$ where $\Delta$ is
the BCS
``bulk gap'' and $d$ is the mean level spacing for the single electron
energies.
For nanoscale systems, this ratio is of the order of unity,
in
seeming contradiction with the experimental results discussed above.
The explanation for this is that the mean-field approach is
inappropriate
in this instance, as previously indicated.

\section{Quantum inverse scattering method}

The essential motivation for the quantum inverse scattering method is the
construction of a family of commuting matrices, known as transfer
matrices. 
That is, we wish to construct an operator $t(u)$, where $u\in\mathbb{C}$
is called the spectral parameter, acting on some vector space,
which represents the Hilbert space of physical states. Further we
seek that  
\beq [t(u),\,t(v)]=0~~~\forall \,u,v\,\in{\mathbb{C}}. \label{ctm} \eeq
There are two significant consequences of (\ref{ctm}). The first is that
$t(u)$ may be diagonalised independently of $u$, that is the
eigenvectors
of $t(u)$ do not depend on $u$. This is the feature which makes the
Bethe ansatz approach viable. Secondly, $t(u)$ commutes with all
of its derivatives, or more formally, taking the series expansion
$$t(u)=\sum_{k=-\infty}^{\infty}\cC_ku^k $$
it follows that
$$[\cC_k,\,\cC_j]=0 ~~~~\forall\,k,\,j.$$
Thus for any Hamiltonian which is expressible as a function of the
operators $\cC_k$ only, each $\cC_k$ corresponds to
an operator representing a constant of the motion, since it will commute
with the Hamiltonian.
When the number of independent conserved quantities is equal to the number of
degrees of freedom of the system, the model is said to be integrable.

Let $V$ denote some fixed vector space of finite-dimension $n$. 
The theory of exactly solvable quantum systems in this 
setting begins with an invertible operator, depending on the
spectral parameter $u$,
$$R(u)\in {\rm End} (V\otimes V)$$ 
called the $R$-matrix. Here ``End'' refers to the
space of endomorphisms (square matrices), 
so $R(u)$ is effectively an $n^2\times n^2$
matrix whose entries are scalar functions of $u$.  
From the $R$-matrix we define the  
Yang-Baxter algebra, denoted $Y$, 
which is generated by the monodromy matrix $T(u)$,  
whose entries are elements of $Y$  
\beq
R_{12}(u-v) T_1(u) T_2(v)=
T_2(v) T_1(u)R_{12}(u-v). \label{yba}
\eeq
The above equation acts in the three-fold space ${\rm End}(V\otimes V)
\otimes Y$ and the subscripts refer to the components of ${\rm
End}(V\otimes V)$. In terms of the elementary matrices $e^i_j$, which have
1 in the $(i,j)$ position and zeros elsewhere, we may write 
\bea &&R(u)=\sum_{i,j,k,l=1}^n R^{ik}_{jl}(u)\,e^i_j\otimes e^k_l, \nn \\
&&T(u)=\sum_{i,j=1}^ne^i_j\otimes T^i_j(u). \nn \eea 
Then 
\bea &&R_{12}(u)=\sum_{i,j,k,l=1}^n R^{ik}_{jl}(u)\,e^i_j\otimes e^k_l\otimes
I, \nn \\ 
&&T_1(u)=\sum_{i,j=1}^n e^i_j\otimes I \otimes T^i_j(u), \nn \\
&&T_2(u)=\sum_{i,j=1}^n I\otimes e^i_j\otimes T^i_j(u) \nn \eea 
where $I$ is the identity operator. In component form we may write  
\beq \sum_{j,l=1}^n R_{jl}^{ik}(u-v)T_p^j(u)T_q^l(v)
=\sum_{j,l=1}^nT_j^k(v)T_l^i(u)R_{pq}^{lj}(u-v) \label{summ} \eeq 
so the $R^{ik}_{jl}(u)$ give the structure constants of the
algebra. Note that $Y$ is actually an infinite-dimensional algebra, a
basis for which $\{T^i_j[k]\}$ is obtained by taking the series expansions 
$$T^i_j(u)=\sum_{k=-\infty}^{\infty} u^k T^i_j[k]. $$    

Imposing that $Y$ is an associative algebra leads, through repeated use
of (\ref{yba}), to the following
equation in ${\rm End}\,(V\otimes V\otimes V)\otimes Y$    
\bea 
&&T_{1}(u)T_{2}(v)T_{3}(w)\nn \\
&&~~=\l(T_{1}(u)T_{2}(v)\r)T_{3}(w)\nn \\
&&~~=R^{-1}_{12}(u-v)\l(T_{2}(v)T_{1}(u)\r)T_{3}(w)R_{12}(u-v) \nn \\
&&~~=\cdots\cdots\cdots
=R^{-1}_{12}(u-v)R^{-1}_{13}(u-w)R^{-1}_{23}(v-w)T_{3}(w)T_{2}(v)
T_{1}(u) \nn \\
&&~~~~~~~~~~~~~~~~~~~~~~~~~~~~~~
\times R_{23}(v-w)R_{13}(u-w)R_{12}(u-v). \label{assoc1} \eea 
Here $R_{jk}(u)$ denotes the matrix in ${\rm End}
(V \otimes V\otimes V)$ acting non-trivially on the
$j$-th and $k$-th spaces and as the identity on the remaining space.
In a similar way one may deduce that 
\bea 
&&T_{1}(u)T_{2}(v)T_{3}(w)\nn \\
&&~~=T_{1}(u)\l(T_{2}(v)T_{3}(w)\r)\nn \\
&&~~=R^{-1}_{23}(v-w)R^{-1}_{13}(u-w)R^{-1}_{12}(u-v)T_{3}(w)T_{2}(v)
T_{1}(u) \nn \\   
&&~~~~~~~~~~~~~~~~\times R_{12}(u-v)R_{13}(u-w)R_{23}(v-w). \label{assoc2} \eea 
A sufficient condition for (\ref{assoc1}) and (\ref{assoc2}) to be
equivalent is that the $R$-matrix
satisfies the {\it Yang-Baxter equation} 
acting in ${\rm End}(V\otimes V\otimes V)$
\beq
R _{12} (u-v)  R _{13} (u-w)  R _{23} (v-w) =
R _{23} (v-w)  R _{13}(u-w)  R _{12} (u-v).
\label{ybe} \eeq
The above shows that in this algebraic setting the Yang-Baxter equation
arises as a natural way to impose associativity of the Yang-Baxter
algebra $Y$. It also appears in many other contexts, such as 
classical two-dimensional statistical mechanics \cite{baxter}, knot theory
\cite{knot1,knot2} and scattering theory \cite{zam}. 

\begin{figure}[h]
\includegraphics[width=\textwidth]{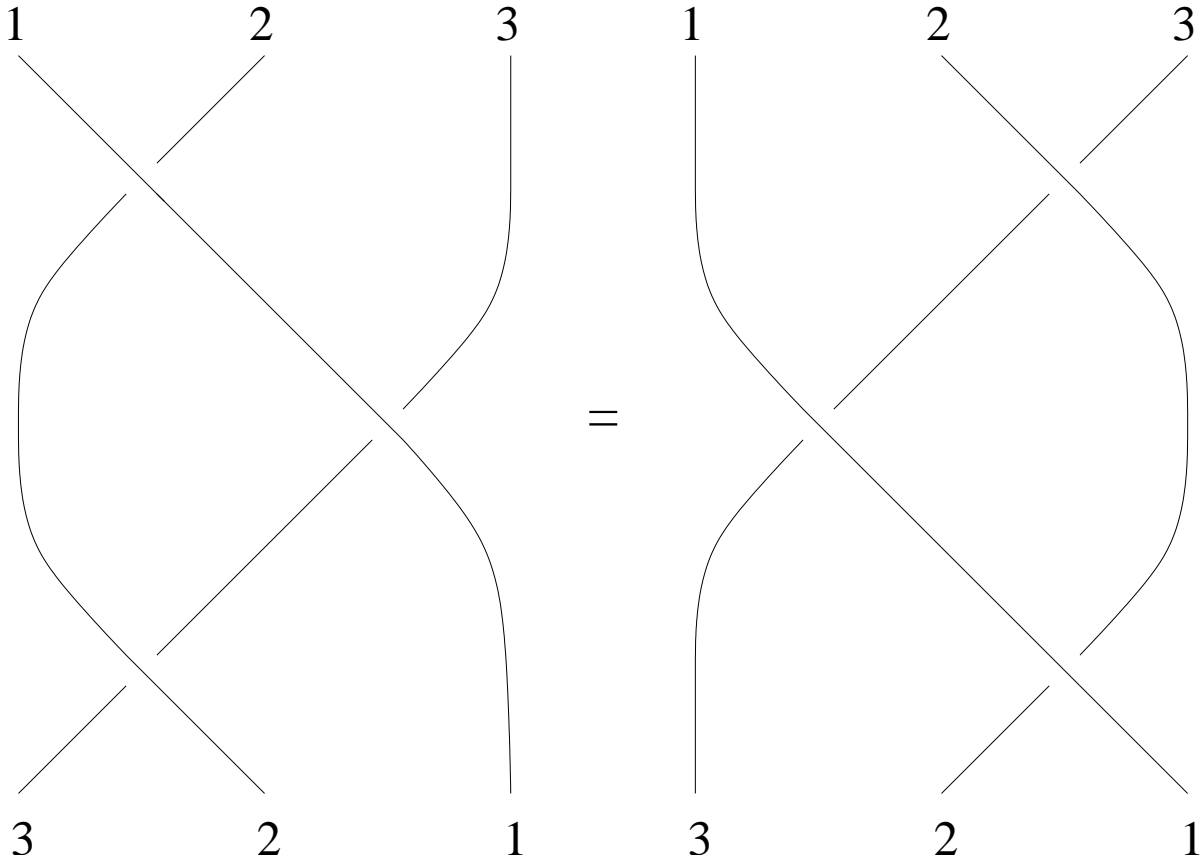}
\caption{
Graphical representation of the Yang-Baxter equation. In the context of
knot theory, $R_{ij}(\infty)$ 
denotes the crossing of string $i$ over string $j$.
The Yang-Baxter equation states that the above two combinations of crossings
are topologically equivalent. In scattering theory the matrix elements of
$R_{ij}(u-v)$ give the amplitudes for the two-body scattering
of particles labelled $i$ and $j$, with rapidity variables $u$ and $v$
respectively. The fact that the scattering depends only on the
difference $u-v$ is a consequence of Lorentz invariance. 
The Yang-Baxter equation is a
statement of equivalence for the two factorisations of three-body
scattering in terms of two-body scattering. In classical two-dimensional
statistical mechanics the matrix elements of $R_{ij}(u)$ give the 
allowed vertex weights at the lattice site labelled by ($i,\, j$). In
this instance $u$ can be parameterised in terms of the energy levels and
temperature. The
Yang-Baxter equation ensures the commutativity of the row-to-row
transfer matrix, from which the partition function is constructed.
}
\label{fig1}
\end{figure}

Here, we will only concern ourselves with the $gl(2)$ invariant $R$-matrix,
which has the form \cite{mcguire,yang}  
\bea
R(u)& = & \frac{1}{u+\eta}({u.I\otimes I+\eta P})\nn \\
&=& \left ( \begin {array} {cccc}
1&0&0&0\\
0&b(u)&c(u)&0\\
0&c(u)&b(u)&0\\
0&0&0&1\\
\end {array} \right ),
\label{rm} \eea
with $b(u)=u/(u+\eta)$,   
$c(u)=\eta/(u+\eta)$ and $\eta$ is an arbitrary complex parameter.
Above, $P$ is the permutation operator which satisfies 
$$P(x\otimes y)=y\otimes x~~~~\forall\,x,\,y\in\,V.$$ 
The $R$-matrix is $gl(2)$ invariant in that 
\beq [R(u),\,\g\otimes \g]=0 \label{gl2} \eeq 
where $\g$ is any $2\times 2$ matrix. 

For this case the Yang-Baxter algebra, denoted $Y[gl(2)]$ has four elements  
\beq  
T(u)  = \left ( \begin {array} {cc}
T^1_1(u)&T^1_2(u)\\
T^2_1(u)&T^2_2(u)
\end {array} \right ). \label{mono}
\eeq
For clarity and convenience we adopt the notation 
$$A(u)=T^1_1(u),~~~B(u)=T^1_2(u),~~~ C(u)=T^2_1(u),~~~ D(u)=T^2_2(u). $$
The full set of algebraic relations governed by (\ref{yba}) are 
\bea
[A(u),\, A(v)] &=& [D(u),\, D(v)] = 0, \nn \\
\l [B(u),\, B(v)\r ] &=& [C(u), \,C(v)] = 0, \no\\
\l [A(u),\,D(v)\r ]&=&\frac{\eta}{u-v}\left(C(v)B(u)-C(u)B(v)\right) \nn \\
\phantom{[A(u),\,D(v)]}&=&\frac{\eta}{u-v}\left(B(u)C(v)-B(v)C(u)\right),\nn
\\
A(u)B(v) &=& 
\frac {u-v-\eta}{u-v} B(v) A(u)+\frac {\eta}{u-v}
B(u) A(v), \no\\
 A(u)C(v) &=& \frac {u-v+\eta}{u-v} C(v) A(u)-\frac {\eta}{u-v}
	C(u) A(v), \no\\
 D (u) B(v) &=& 
\frac {u-v+\eta}{u-v} B(v) D(u) -\frac
{\eta}{u-v}
B(u) D(v), \nn \\
 D (u) C(v)&=& 
	\frac {u-v-\eta}{u-v} C(v) D(u) +\frac {\eta}{u-v}
	C(u) D(v), \nn \\
 B(u)A(v) &=& \frac {u-v-\eta}{u-v} A(v) B(u)+\frac {\eta}{u-v}
			A(u) B(v), \no\\
 B(u)D(v) &=& \frac {u-v+\eta}{u-v} D(v) B(u) -\frac {\eta}{u-v}
		D(u) B(v), \nn \\
 C(u)A(v) &=& \frac {u-v+\eta}{u-v} A(v) C(u)-\frac {\eta}{u-v}
						A(u) C(v), \no\\
 C(u)D(v) &=& \frac {u-v-\eta}{u-v} D(v) C(u) +\frac
         {\eta}{u-v} D(u) C(v), \nn \\
\l [B(u),\,C(v)\r ]&=&\frac{\eta}{u-v}\left(A(u)D(v)-A(v)D(u)\right) \nn \\
\phantom{[B(u),\,C(v)]}&=&\frac{\eta}{u-v}\left(D(v)A(u)-D(u)A(v)\right). 
\label{rels} \eea

Next suppose that we have a realisation of $Y[gl(2)]$ acting on some vector
space $W$, which we denote $\pi:Y[gl(2)]\rightarrow 
{\rm End}\,W$.  
It is usual to refer to $V$ as the auxiliary space and $W$ as the
physical space. Note that as $Y[gl(2)]$ is a quadratic algebra,  any
realisation can be multiplied by  an overall scaling factor and still
satisfy the relations (\ref{rels}). 
For later convenience we set 
$$L(u)=\pi\left( T(u)\right)\in\,{\rm End}\,(V\otimes W) $$ 
which we refer to as an $L$-operator.   
Defining the transfer matrix through 
\beq 
t(u) = \pi \l({\rm tr}\, (T(u))\right)=\pi\left(A(u)+D(u)\right)
\in\,{\rm End}\, W\label{tm} \eeq  
it follows from (\ref{yba}) that the 
transfer matrices commute for different values of the spectral parameter;
viz. equation (\ref{ctm}) is satisfied.  

An important property of the Yang-Baxter algebra is that it has a 
co-multiplication structure which allows us to build tensor product 
realisations. In particular, given two $L$-operators 
$L^U \in\,{\rm End}\,(V\otimes U)$ 
and $L^W \in\,{\rm End}\,(V\otimes W)$, 
then $L=L^UL^W\in{\rm End}\,(V\otimes U\otimes W)$ 
is also an $L$-operator as can be seen from 
\bea R_{12}(u-v)L_1(u)L_2(v)&=&R_{12}(u-v)L^U_1(u)L^W_1(u)L^U_2(v)L^W_2(v)\nn \\
&=&R_{12}(u-v)L^U_1(u)L^U_2(v)L^W_1(u)L^W_2(v) \nn \\
&=&L^U_2(v)L^U_1(u)R_{12}(u-v)L^W_1(u)L^W_2(v) \nn \\
&=&L^U_2(v)L^U_1(u)L^W_2(v)L^W_1(u)R_{12}(u-v) \nn \\
&=&L^U_2(v)L^W_2(v)L^U_1(u)L^W_1(u)R_{12}(u-v) \nn \\
&=&L_2(v)L_1(u)R_{12}(u-v). \nn \eea 
Furthermore, if $L(u)$ is an $L$-operator then so is $L(u+\alpha)$ for any 
$\alpha$, since the $R$-matrix depends only on the difference of the 
spectral parameters. This property will prove important in all
constructions below. 

\subsection{The quasi-classical limit} 

The $R$-matrix (\ref{rm}) has the property
$$\lim_{\eta\rightarrow 0} R(u)=I\otimes I$$ 
which is known as the quasi-classical property. For any such $R$-matrix
it is appropriate to write 
\bea R(u)&=&I\otimes I +\eta \R(u) +o(\eta^2) \nn \\
T^i_j(u)&=&\delta^i_j I +\eta \T^i_j(u)+ o(\eta^2) \nn \eea 
and substitute into (\ref{summ}). Equating the second order terms in $\eta$
yields the following relations  
\bea [\T^i_p(u),\,\T^k_q(v)]
&=&\sum_{j=1}^n\l(\R^{jk}_{pq}(u-v)\T^i_j(u)-\R^{ik}_{jq}(u-v)\T^j_p(u)\r.
\nn \\ 
&&~~~~~~~\l.+\R^{ij}_{pq}(u-v)\T^k_j(v)-\R^{ik}_{pj}(u-v)\T^j_q(v)\r)\nn 
\eea 
which we take to be the defining relations for the algebra denoted $\Y$,
to be called the Gaudin algebra. Gaudin used the quasi-classical
limit to define classes of integrable spin chain Hamiltonians with
long-range interactions \cite{gaudin1,gaudin2}. The algebraic approach
which we follow here is due to Sklyanin \cite{sklyanin2,sklyanin1}. 
Observe that in the quasi-classical limit $\Y$
is an infinite-dimensional 
Lie algebra in contrast to the quadratic algebra structure of $Y$.
Any realisation of $Y$ which admits the quasi-classical limit 
provides a realisation of $\Y$. 

For the case of the $gl(2)$ invariant $R$-matrix (\ref{rm}) let us write 
\bea A(u)&=& I+\eta \A(u) +o(\eta^2), \nn \\
B(u)&=&\eta \B(u)+o(\eta^2), \nn \\
C(u)&=&\eta \C(u)+o(\eta^2), \nn \\
D(u)&=&I +\eta \D(u)+o(\eta^2). \nn \eea 
From (\ref{rels}) we determine that the full relations for the algebra 
$\Y[gl(2)]$ are 
\bea [\A(u),\,\A(v)]&=& [\B(u),\,\B(v)]=0,\nn \\
\l[\C(u),\,\C(v)\r]&=& [\D(u),\,\D(v)]=0, \nn \\
\l[\A(u),\,\D(v)\r]&=&0, \nn \\
\l[\B(u),\,\C(v)\r]&=& \frac{\A(u)-\A(v)+\D(v)-\D(u)}{u-v}, \nn \\
\l[\A(u),\,\B(v)\r]&=&\frac{\B(u)-\B(v)}{u-v},~~~~
\l[\A(u),\,\C(v)\r]=\frac{\C(v)-\C(u)}{u-v}, \nn \\
\l[\D(u),\,\B(v)\r]&=&\frac{\B(v)-\B(u)}{u-v},~~~~
\l[\D(u),\,\C(v)\r]=\frac{\C(u)-\C(v)}{u-v}.  \nn \eea 

\subsection{Examples of realisation of the Yang-Baxter algebra}

In order to construct a specific model, we must address the question of 
determining a realisation of the Yang-Baxter algebra. Here we present 
several examples.  The first realisation
comes from the $R$-matrix itself, since it is apparent by setting $w=0$
in (\ref{ybe}) 
that we 
can make the identification $L(u)=R(u)$ such that a realisation of 
(\ref{yba}) is obtained. This realisation satisfies the
quasi-classical property, and 
is that  used in the construction of the Heisenberg model 
\cite{fst,ks,takhtajan,korepin,faddeev}. 
A second realisation is given by $L(u)=\g$ ($c$-number realisation), 
where $\g$ is an arbitrary $2\times 
2$ matrix whose entries do not depend on $u$ (although can depend on
$\eta$). 
This follows from the fact that (\ref{gl2}) holds for any such $\g$. 
 
There is a realisation 
in terms of canonical boson operators $a,\,a^{\dagger}$ with the relation
$[a,\,a^{\dagger}]=1$ which reads
\cite{kt}
\beq 
L^{a}(u) = \left ( \begin {array} {cc}
(1+\eta u)I+\eta^2 N &\eta a\\
\eta a^\dagger & I\\
\end {array} \right )
\label{lb} \eeq 
where $N=a^{\dagger}a$. 
There also exists a realisation in terms of the $su(2)$ Lie algebra 
 with generators $S^z$ and $S^{\pm}$ \cite{fst,ks,takhtajan,korepin,faddeev},
\beq 
 L^S(u) = \frac{1}{u}\left ( \begin {array} {cc}
 uI+{\eta} S^z &{\eta} S^-\\
 {\eta} S^+& uI-{\eta} S^z\\
 \end {array} \right ),
\label{ls} \eeq  
 subject to the commutation relations 
\beq [S^z, S^{\pm}] = \pm S^\pm, ~~~~[S^+,S^-] =2 S^z.\label{su2} \eeq 
When the $su(2)$ algebra
takes the spin 1/2 representation the resulting $L$-operator is equivalent
to that given by the $R$-matrix.  
 Another is realised in terms of the  $su(1,1)$
 generators $K^z$ and $K^{\pm}$ \cite{jurco,rybin},
\beq 
 L^K(u) = \frac{1}{u} \left ( \begin {array} {cc}
 uI+\eta K^z &{\eta} K^-\\
 -{\eta} K^+& uI-{\eta} K^z\\
 \end {array} \right ),
\label{lk} \eeq  
 with the commutation relations 
\beq [K^z, K^{\pm}] = \pm K^\pm, ~~~~[K^+,K^-] =-2 K^z. \label{su11} \eeq  
Each of the realisations $L^a(u),\,L^S(u)$ and $L^K(u)$ satisfy the
quasi-classical property, and thus affords a realisation of the Gaudin
algebra.

The discerning reader may notice that $L^S(u)$ and $L^K(u)$ are in fact
equivalent, which results from the Lie algebra homomorphism
$\Upsilon:\, su(2)\rightarrow
su(1,1)$ defined by 
$$\Upsilon(S^z)=K^z,~~~~\Upsilon(S^+)=-K^+,~~~~\Upsilon(S^-)=K^-    $$  
such that (\ref{su2}) is mapped to (\ref{su11}). 
For convenience we make the distinction between these two $L$-operators 
as the transformation
$\Upsilon$ is non-unitary. This permits us to avoid the use of non-unitary
realisations of the $su(2)$ algebra below. 
(Although, as will be seen, the realisations of the Yang-Baxter algebra
may not be unitary.) 

\subsection{${\mathbb{Z}}$-graded structure and $\Z$-graded realisations 
of the Yang-Baxter algebra}

The Yang-Baxter algebra $Y[gl(2)]$ carries a
$\Z$-graded structure that can be exploited in the construction of models
of Bose-Einstein condensates, as explained in \cite{zlgm}, which we now
recount. 
We introduce an auxiliary operator $Z$, called the {\it grading
operator}, satisfying the relations
\beq  [Z,\,X(u)]=p\{X(u)\}.X(u), \label{grading} \eeq
where $X=A,\,B,\,C$ or $D$ and 
$$p\{A(u)\}=p\{D(u)\}=0,~~\,p\{B(u)\}=1,~~\, 
p\{C(u)\}=-1.$$
We call $p\{X(u)\}\,\in\,{\mathbb{Z}}$
the {\it gradation} of $X(u)$, and extend the gradation operation
to the entire algebra
by  the requirement
$$p\{\theta.\phi\}=p\{\theta\}+p\{\phi\}~~~~\forall \,\theta,\,\phi\,\in
\A . $$
This definition for the grading operator is consistent with the defining
relations (\ref{rels}).

Let us now define a class of realisations of $Y[gl(2)]$ 
which we call ${\mathbb Z}$-graded
realisations.  We say that  a vector space $W$, equipped
with an endomorphism $z$,
is a $\Z$-graded vector space, denoted $(W,z)$,
if it admits a
decomposition into subspaces
$$W=\bigoplus_{j=-\infty}^\infty W_j$$
such that
$$zW_j=j.W_j, ~~~~~~j\,\in\, \Z. $$
Note that some of the $W_j$ may be trivial subspaces.
Formally, the grading operator can be used to define the following
projection operators
\beq P_j=\prod_{\stackrel{k=-\infty}{k\neq j}}^\infty
\frac{(z-kI)}{(j-k)}
\label{proj} \eeq
such that
$$P_kP_j=\delta_{kj} P_j,~~~~P_kW_j=\delta_{kj}W_k. $$
We say that a $\Z$-graded vector space
$$W'=\bigoplus_{j=-\infty}^\infty W'_j$$
is {\it equivalent} to $W$ if
for some $k\,\in\,\Z$ there exist a vector space isomorphism between
$W'_j$ and $W_{j+k}$ for all $j$.
This terminology is motivated by the fact that for a given $(W,z)$ one
can always generate another $\Z$-graded space $(W',z')$ through
the mappings
$W'_j\rightarrow W_{j+k},\,z'\rightarrow z-kI$ for any $k\,\in\,\Z$.

For a given $\Z$-graded $W$ we say that
$\pi :\,Y[gl(2)]\rightarrow  {\rm End}\, W$ provides a $\Z$-graded
realisation of $Y[gl(2)]$ if $\pi(Z)=z$ and 
the relations (\ref{rels},\,\ref{grading})
are preserved. In such a case  we can write 
$$\pi(X(u))=\sum_{j=-\infty}^\infty X(u,j)$$
and the matrices
$X(u,j)$ satisfy
$$X(u,j)W_k=0 ~~~~~{\rm for}\,\, j\neq k. $$
More specifically, this means that for $\left|\psi_j\right>\,\in\,W_j$
we have
$$\pi(X(u)Y(v))\left|\psi_j\right>
=X(u,j+p\{Y(u)\})Y(v,j)\left|\psi_j\right>. $$

In view of the equivalence of $\Z$-graded vector spaces defined above,
there can also exist equivalent realisations. We can define
a realisation $\pi'$ equivalent to $\pi$
by specifying   some $k\,\in\,\Z$ such that
$$\pi'(Z)= \pi(Z-kI)  $$
and for
$$\pi'(X(u))=\sum_{j=-\infty}^\infty X'(u,j)$$
the matrices $X'(u,j)$ are defined by
$$X'(u,j)= X(u,j+k)~~~~\forall \,j\,\in\,\Z. $$

For the $\Z$-graded case  we may
express the 
transfer matrix as
$$t(u)=\sum_{j=-\infty}^\infty t(u,j) $$
such that
$$t(u,j)W_k=0~~~~~~{\rm for}\,\, j\neq k$$
and $$[t(u,j),\,t(v,k)]=0~~~~~\forall\,j,\,k\in\,\Z,\,u,\,v\in\,
{\mathbb C}. $$
Since $p\{t(u)\}=0$, the diagonalisation of $t(u)$
is thus reduced to the diagonalisation
of each of the matrices $t(u,j)$ on the
$\Z$-graded component $W_j$,
where we have
$$[t(u,j),\,t(v,j)]=0~~~~~\forall\,u,\,v\in\,{\mathbb C}. $$
We may restrict our attention to the
case of $t(u,0)$, as each $t(u,j)$ is
equivalent to some
$t'(u,0)$ through the use of equivalent
realisations as introduced
above.

\subsection{Examples of $\Z$-graded realisations} 

Next we give  two non-trivial $\Z$-graded realisations of the algebra
$Y[gl(2)]$. One is expressible in terms of two Heisenberg algebras with
generators $a_i,\,a_i^{\d},\,i=1,\,2$ and reads
$X(u,j)=\tilde{X}(u,j)P_j$ with
\bea
\tilde{A}(u,j)) &=& u^2 + \eta u N +\eta ^2 N_1N_2 -\eta
(N_1-N_2)\o(N+jI)\no\\
&&-\o^2(N+jI) +a^\d_2 a_1, \no\\
\tilde{B}(u,j) &=& (u+\o(N+jI)+\eta N_1)a_2+\eta^{-1}a_1, \no\\
\tilde{C}(u,j) &=& a^\d_1 (u-\o(N+jI)+\eta N_2)+\eta^{-1}a^{\d}_2, \no
\\
\tilde{D}(u,j) &=& a_1^{\d}a_2+\eta^{-2}, \nn \\
Z&=&k.I-N. \label{zg1} \eea
Above, $k$ is an arbitrary scalar, 
$P_j$ are the projections defined by (\ref{proj}),
$N_i=a^{\d}_ia_i$,  $N=N_1+N_2$ and $\o(x)$ is an
arbitrary polynomial function of $x$.
The operators act on the Fock space spanned by the basis vectors given
by (\ref{fock}). 
Note that in the case when $\o(x)$ is constant, the above realisation
reduces to that discussed in \cite{enolskii1,enolskii2,lz,flz,zlmx} 
and is factorizable into two local
realisations of the Yang-Baxter algebra expressible in terms of the
two Heisenberg algebras; viz. 
$$L(u)=\eta^{-2}L^a_1(u-\eta^{-1}+\omega)L_2^a(u-\eta^{-1}-\omega).$$ 
It is
important to note that for generic $\o(x)$ no such factorisation exists.

Another $\Z$-graded realisation of the Yang-Baxter algebra
is $X(u,j)=\tilde{X}(u,j)P_j$ with
\bea
\tilde{A}(u,j) &=& -\eta u^2 +
u\left(1-\eta^2 (K^z+N_c) -\eta \o(K^z+N_c+jI)\right)\no\\
&&+\eta K^z -\eta^2 K^z \o(K^z+N_c+jI) -\eta^3 N_cK^z +\eta^2 c K^+,
\no\\
\tilde{B}(u,j) &=& \eta (1-\eta u- \eta \o(K^z+N_c+jI)-\eta^2 N_c)K^-
-\eta c(u-\eta K^z), \no\\
\tilde{C}(u,j) &=& \eta c^\dagger (u+\eta K^z)-\eta K^+, \no \\
\tilde{D}(u,j) &=& u-\eta K^z +\eta^2 c^\dagger K^-, \nn \\
Z&=&k.I-K^z-N_c+\kappa.\label{zg2}
\eea
Above, $k$ is again arbitrary, 
the operators $c,\,c^{\d}$ form a Heisenberg algebra, with
$N_c=c^{\d}c$, and as before the operators $K^z,\,K^+,\,K^-$ 
 satisfy the 
$su(1,1)$ relations (\ref{su11}).  
It is assumed that the $su(1,1)$ operators are realised in terms of an
irreducible representations of lowest weight $\kappa$. 
As in the previous example, $\o(x)$ is an arbitrary polynomial function
of $x$ and
the above realisation is factorisable only in the case when $\o(x)$ is
constant. In this instance we have 
$$L(u)=u\g L^a(u-\eta^{-1}+\omega)L^K(u) $$  
where $\g={\rm diag}(-1,\,1)$. 

\section{Algebraic Bethe ansatz method of solution}

A key step in successfully applying the algebraic Bethe ansatz approach
is finding
a suitable pseudovacuum state, $\left|\chi\right>$, which has the properties
\bea A(u)\left|0\right>&=&a(u)\left|\chi\right>, \nn \\
B(u)\left|0\right>&=&0, \nn \\ 
C(u)\left|0\right>&\neq & 0, \nn \\
D(u)\left|0\right>&=&d(u)\left|\chi\right> \nn \eea 
where $a(u)$ and $d(u)$ are scalar functions. 
Note that for ease of notation throughout we will omit the symbol $\pi$
denoting the realisation of $Y[gl(2)]$. 
Next choose the Bethe state 
\beq \left|\vec v\right>\equiv\left|v_1,...,v_M\right>
= \prod ^M_{i =1} C(v_i) \left|\chi \right>. \label{state} \eeq 
Note that because $[C(u),\,C(v)]=0$, the ordering is not important in
the product of (\ref{state}). 
The approach of the algebraic Bethe ansatz is to
use the relations (\ref{rels}) to determine the action of $t(u)$ on
$\left|\vec v\right>$. 
First let us consider the action of $A(u)$ on $\left|\vec v\right>$;
viz.
$$A(u)\left|\vec v\right>=A(u)C(u_i)\left|\vec v_i\right> $$ 
where 
$$  \left|\vec v_i\right>\equiv\left|v_1,\cdots,v_{i-1},v_{i+1},
\cdots,v_M\right>
.$$ 
Now
\bea A(u)\l|\vec v\r>&=&\frac{u-v_i+\eta}{u-v_i}C(v_i) A(u)
\l|\vec v_i\r>-\frac{\eta}{u-v_i}C(u)A(v_i)\l|\vec v_i\r> \nn
\\
&=&\frac{u-v_i+\eta}{u-v_i}C(v_i)A(u)C(v_j)\l|\vec v_{ij}\r>
-\frac{\eta}{u-v_i}C(u)A(v_i)C(v_j)\l|\vec v_{ij}\r> \nn
\\
&=&\l(\frac{u-v_i+\eta}{u-v_i}\r)\l(\frac{u-v_j+\eta}{u-v_j}\r)
C(v_i)C(v_j)
A(u)\l|\vec v_{ij}\r> \nn \\
&&-\l(\frac{u-v_i+\eta}{u-v_i}\r)\l(\frac{ \eta}{
u-v_j}\r)C(v_i)C(u)A(v_j)\l|\vec v_{ij}\r> \nn \\
&&-\l(\frac{\eta}{u-v_i}\r)\l(\frac{v_i-v_j+\eta}{v_i-v_j}\r)C(u)C(v_j)
A(v_i)\l|\vec v_{ij}\r> \nn \\
&&+\l(\frac{\eta}{u-v_i}\r)\l(\frac{\eta}{v_i-v_j}
\r)C(u)C(v_i)A(v_j)\l|\vec v_{ij}\r>, \label{acom} \eea
where, 
$$\l|\vec v_{ij}\r>\equiv \l|v_1,\cdots,v_{i-1},v_{i+1},\cdots,v_{j-1},v_{j+1},
\cdots,v_M\r>.$$ 
Proceeding further we find the general form
\bea A(u)\l|\vec v\r>&=&\l(\prod_{i=1}^M\frac{u-v_i+\eta}{u-v_i}\r)
\l(\prod_{i=1}^M C(v_i)\r)A(u)\l|\chi\r> \nn \\
&&-\frac{\eta}{u-v_i}\l(\prod^M_{j\neq
i}\frac{v_i-v_j+\eta}{v_i-v_j}\r)
C(u)\l(\prod_{k\neq i}^M C(v_k)\r)A(v_i)\l|\chi\r> \nn \\
&&+~\mathrm{ other~ linearly~ independent~ terms} \nn \\
&=&a(u)\l(\prod_{i=1}^M\frac{u-v_i+\eta}{u-v_i}\r)\l|\vec v\r> \nn \\
&&-\frac{\eta a(v_i)}{u-v_i}\l(\prod^M_{j\neq
i}\frac{v_i-v_j+\eta}{v_i-v_j}\r)C(u)\l|\vec v_i\r> \nn
\\
&&+~\mathrm{ other~ linearly~ independent~ terms.} \nn \ea
Above, each of the other linearly independent terms is a vector of the
form $C(u)\l|\vec v_j\r>,\,j\neq i$, multiplied by some scalar. There are
no other possibilities. To determine what the co-efficients are we note
that the above equation is valid for any choice of $i$.
Hence  we conclude that
\bea A(u)\l|\vec v\r>&=&a(u)\l(\prod_{i=1}^M\frac{u-v_i+\eta}{u-v_i}
\r)\l|\vec v\r> \nn \\
&&-\sum_{i=1}^M\frac{\eta a(v_i)}{u-v_i}\l(\prod^M_{j\neq i}\frac{
v_i-v_j+\eta}{v_i-v_j}\r)C(u)\l|\vec v_i\r>. \nn \ea
The co-efficients of the terms $C(u)\l|\vec v_i\r>$ are called
{\it unwanted terms} for reasons that will soon become apparent.

We now perform the same procedure for $D(u)$;
\bea D(u)\l|\vec v\r>&=& D(u)C(v_i)\l|\vec v_i\r> \nn
\\
&=&\l(\frac{u-v_i-\eta}{u-v_i}\r)C(v_i)D(u)\l|\vec v_i\r>+
\frac{\eta}{u-v_i}C(u)D(v_i)\l|\vec v_i\r> \nn \\
&=&d(u)\l(\prod_{i=1}^M\frac{u-v_i-\eta}{u-v_i}\r)\l|\vec v\r> \nn
\\
&&+\sum_{i=1}^M\frac{\eta d(v_i)}{u-v_i}\l(\prod^M_{j\neq i}
\frac{v_i-v_j-\eta}
{ v_i-v_j}\r)C(u)\l|\vec v_i\r>. \label{d} \eea

The final result for the action of the transfer matrix is 
\bea  
t(u) \left|\vec v\right> 
&=& \l(A(u)+D(u)\r)\l|\vec v\r> \nn \\
&=&\La (u,\,\vec v) 
\left|\vec v\right>  \nn \\
&&~~-\sum_i^M\frac{\eta a(v_i)}{u-v_i}\l(\prod_{j\neq i}^{M}
\frac{v_i-v_j+\eta}{v_i-v_j}
\right)C(u)\left|\vec v_i\right> \nn \\ 
&&~~+\sum_{i=1}^M\frac{\eta d(v_i)}{u-v_i}\l(\prod_{j\neq i}^{M}
\frac{v_i-v_j-\eta}{v_i-v_j}
\right)C(u)\left|\vec v_i\right> \label{osba}  
\eea 
where
\beq 
\La(u,\,\vec v) = a(u) \prod ^M_{i=1} 
\frac {u-v_i+\eta}
{u-v_i}+
d(u) \prod ^M_{i=1} \frac {u-v_i-\eta}
{u-v_i}.  
\label{tme} \eeq 
The above shows that $\left|\vec v\right>$ becomes an eigenstate of the
transfer matrix with eigenvalue (\ref{tme}) whenever
the unwanted terms cancel. This occurs when the {\it Bethe ansatz equations} 
\beq
\frac{a(v_i)}{d(v_i)}=
\prod ^M_{j \neq i}\frac {v_i -v_j - \eta}{v_i -v_j +\eta},
~~~~~~~i=1,...,M.\label{bae} \eeq
are satisfied. 
Throughout we adopt the notation
$$\{v_i\}\equiv \{v_1,v_2,\cdots,v_M\}$$ 
for such a solution.

Note that in the derivation of the Bethe ansatz equations
 it is required that $v_i\neq v_j\,
\,\forall\,i,\,j.$ 
This is a result of the Pauli Principle for Bethe 
ansatz solvable models, as developed in \cite{ik} for the
one-dimensional Bose gas with delta-function interactions. 
For generic functions $a(u),\,d(u)$, 
essentially the same argument as \cite{ik} can be applied to draw the
same conclusion. To give an indication why this is the case, consider
(\ref{acom}) in the limit $v_i\rightarrow v_j$
\bea A(u)\l|\vec v\r>
&=&\l(\frac{u-v_i+\eta}{u-v_i}\r)^2
C(v_i)^2
A(u)\l|\vec v_{ij}\r> \nn \\
&&-\l(\frac{u-v_i+\eta}{u-v_i}\r)\l(\frac{ \eta}{
u-v_i}\r)C(v_i)C(u)A(v_i)\l|\vec v_{ij}\r> \nn \\
&&+\l(\frac{\eta^2}{u-v_i}\r)
C(u)\l[\frac{\rmd}{\rmd w}C(v_i).A(v_i)-C(v_i).\frac{\rmd}{\rmd w}A(v_i)\r]
\l|\vec v_{ij}\r>. \nn \eea
This equation shows that new types of unwanted terms occur which depend
on the derivatives of the elements of $Y[gl(2)]$, and this leads to
an overdetermined system of equations which do not admit a solution. Another
viewpoint is to note that (up to an overall scaling factor, and 
under suitable  assumptions for the forms of
$a(u)$ and $d(u)$) the eigenvalues $\Lambda(u,\vec v)$ are analytic 
functions of $u$. Assuming that the poles in (\ref{tme}) are simple then
the Bethe ansatz equations (\ref{bae}) are equivalent to the statement
that the residue vanishes at each pole; i.e., 
$$\lim_{u\rightarrow v_i}(u-v_i)\Lambda(u,\vec v)=0$$ 
leads to (\ref{bae}). For non-simple poles however
there are  additional Bethe ansatz equations, which 
cannot be satisfied. 
To illustrate this, we consider the simplest case where, for all 
values of $\eta$, all poles are
simple except for one, say at $u=v_j$, which is second order. 
In such a case, let us write 
$$\Lambda(u,\vec v)=a(u)\frac{(u-v_j+\eta)^2}{(u-v_j)^2}
\prod_{k\neq j}^M\frac{(u-v_k+\eta)}{(u-v_k)}+d(u)\frac{(u-v_j-\eta)^2}
{(u-v_j)^2}\prod_{k\neq j}^M\frac{(u-v_k-\eta)}{(u-v_k)}. $$  
Analyticity of $\Lambda(u,\vec v)$ requires 
$$ 0=\lim_{u\rightarrow v_j}(u-v_j)^2\Lambda(u,\vec v) 
$$  
leading to the Bethe ansatz equations 
$$
\frac{a(v_j)}{d(v_j)}=
-\prod ^M_{k \neq j}\frac {v_j -v_k - \eta}{v_j -v_k +\eta}
. $$ 
In particular note that 
\beq \lim_{\eta\rightarrow 0}\frac{a(v_j)}{d(v_j)}=-1. \label{contra} \eeq
Furthermore 
\bea 0&=&\lim_{u\rightarrow v_j}(u-v_j)\Lambda(u,\vec v) \nn \\
&=&\lim_{u\rightarrow v_j}\frac{\rmd}{\rmd u}
\l[(u-v_j)^2\Lambda(u,\vec v)\r] \nn
\\
&=&\lim_{u\rightarrow v_j}\l(a'(u)(u-v_j+\eta)^2\prod_{k\neq j}^M(u-v_k+\eta)
+ 2a(u)\prod_{k=1}^M(u-v_k+\eta)\right.\nn \\
&&~~~~~~~~+ a(u)(u-v_j+\eta)^2\sum_{l\neq j}^M
\prod_{k\neq l,j}^M(u-v_k+\eta) \nn \\ 
&&~~~~~~~~+d'(u)(u-v_j-\eta)^2\prod_{k\neq j}^M(u-v_k-\eta)+ 2 d(u)
\prod_{k=1}^M(u-v_k-\eta) \nn\\  
&&~~~~~~~~\left.+ d(u)(u-v_j-\eta)^2\sum_{l\neq j}^M
\prod_{k\neq l,j}^M(u-v_k-\eta)\right) \nn \\
&=&2\eta a(v_j)\prod_{k\neq j}^M(v_j-v_k+\eta)-2\eta d(v_j)\prod_{k\neq
j}^M(v_j-v_k-\eta) +o(\eta^2). \nn \eea 
The above implies that 
$$\lim_{\eta\rightarrow 0}\frac{a(v_j)}{d(v_j)}=1 $$
in contradiction with (\ref{contra}), proving the claim that one cannot
have a single second order pole. The extension
to more complicated non-simple pole structures, though tedious,
is straightforward.   

Finally, the following identity is useful and is 
easily derived from the Bethe ansatz
equations  
\beq \prod_{i=1}^M a(v_i)=\prod_{i=1}^M d(v_i). \label{identity}\eeq

\subsection{Extension to $\Z$-graded realisations} 

In order to formulate the algebraic Bethe ansatz solution for the class
of $\Z$-graded realisations, we begin with the observation  
from (\ref{rels}) that the following relations hold (amongst others):
\bea
[A(u, j), A(v, j)] &=& [D(u, j), D(v, j)] =0, \no\\
B(u,j) B(v, j-1)& = & B(v,j) B(u, j-1),  \no\\
C(u,j) C(v,j+1) &=& C(v,j) C(u,j+1), \no\\
A(u, j)C(v, j+1) &=& \frac {u-v+\eta}{u-v} C(v,j+1) A(u,j+1)\no\\
&&-\frac {\eta}{u-v} C(u,j+1) A(v,j+1), \no\\
D (u,j) C(v,j+1)& =& \frac {u-v-\eta}{u-v} C(v,j+1) D(u,j+1)\no\\
&&+\frac {\eta}{u-v} C(u,j+1) D(v,j+1), \eea
Again, we  
assume the existence of a pseudovacuum vector
$|\chi \rangle \in W_k$ such that
\bea
A(u,k) |\chi \rangle &=&   a(u,k)|\chi \rangle \nn \\
B(u,k) |\chi \rangle &=& 0 \nn \\
C(u,k) |\chi \rangle &\neq & 0 \nn \\
D(u,k) |\chi \rangle &=&   d(u,k)|\chi \rangle. \nn
\eea
In particular, for the realisation (\ref{zg1}) the Fock vacuum serves as
the pseudovacuum. In the case of (\ref{zg2}) we choose the pseudovacuum to
be the tensor product of the Fock vacuum with the $su(1,1)$ lowest
weight state of weight $\kappa$. 

The above implies that $|\chi \rangle$ is a maximal weight vector
with respect to $Z$. Without loss of generality we can choose $k=M$,
due to the equivalence of realisations discussed earlier,
and look for Bethe states  defined by
\bea
\left|\vec v\r>&=& C(v_1,1)C(v_2,2)\cdots C(v_M,M)|\chi \rangle
\label{estates}.
\eea
It is easy to check that the above Bethe state is symmetric with respect
to interchange of the variables $v_i$, 
a feature which plays a crucial role below.
In particular, this means that we may write
\bea \l|\vec v\r>&=&C(v_i,1)\l|\vec v_i\r> \nn \\
&=&C(v_i,1)C(v_j,2)\l|\vec v_{ij}\r> \nn \\
&=&C(v_j,1)C(v_i,2)\l|\vec v_{ij}\r> \nn \eea 
where now 
\bea \l|\vec v_i\r>&=&C(v_1,2)\cdots C(v_{i-1},i)C(v_{i+1},i+1)\cdots 
C(v_M,M)\l|\chi\r>,  \nn \\
\l|\vec v_{ij}\r>&=&C(v_1,3)\cdots C(v_{i-1},i+1)C(v_{i+1},i+2)\nn \\
&&~~\times \cdots 
C(v_{j-1},j)C(v_{j+1},j+1)\cdots C(v_M,M)\l|\chi\r>. \nn \eea

Acting $A(u,0)$
and $D(u,0)$ on the Bethe state
we have, by following the general procedure detailed above, 
\bea
A(u,0) \l|\vec v\r>&=&  
a (u,M) \l(\prod ^M_{i=1} \frac {u-v_i+\eta}{u-v_i}\r) 
\l|\vec v\r> \no\\
&&~~-\sum^M_{i=1}\frac{\eta a(v_i,M)}{u-v_i}\l(\prod_{j\neq
i}^M \frac{v_i-v_j+\eta}{v_i-v_j}\r)
C(u,1)\l|\vec v_i\r>,\no\\
D(u,0) \l|\vec v\r>&=& 
d(u,M) \l(\prod ^M_{i=1} \frac {u-v_i-\eta}{u-v_i}\r) 
\l|\vec v\r> \no\\
&&~~+\sum^M_{i=1} \frac{\eta d(v_i,M)}{u-v_i}\l(\prod_{j\neq
i}^M \frac{v_i-v_j-\eta}{v_i-v_j}\r)
C(u,1)\l|\vec v_i\r>.\no
\eea
Requiring
$\l|\vec v\r>$  to be an
eigenstate of $t(u,0)$ leads to the Bethe ansatz equations
\beq
\frac {a(v_i,M)}{d(v_i,M)}=
\prod^M_{j \neq i} \frac {v_i-v_j-\eta}{v_i-v_j+\eta},
~~~~~~i=1,\cdots ,M
\label {zgbae} \eeq
and the corresponding eigenvalue of the matrix $t(u,0)$
is
\beq 
\Lambda (u,0,\vec v) =
a(u,M)\prod ^M_{i=1} \frac {u-v_i+\eta}{u-v_i}
+d(u,M)\prod ^M_{i=1} \frac {u-v_i-\eta}{u-v_i} .
\label{zgtme} \eeq

\section{Scalar products of states}

Recall that in the usual algebraic Bethe ansatz for the algebra $Y[gl(2)]$
there is a formula originally due to Slavnov \cite{slavnov} (see also
\cite{korepin,kmt}) for the wavefunction scalar products, which is 
\bea
S(\vec w:\,\vec v) &=& \l\{\vec w|\vec v\r>  \no\\
&=& \l\{\vec v|\vec w\r>  \no\\
&=&\frac {\det F}{\prod^M_{k>l}(v_k-v_l)
\prod^M_{i<j}(w_i-w_j)},
\label{slavnov1} \eea
with the entries of the $M\times M$ matrix $F$ given by
\beq 
F_{ij}=\frac{\eta d(w_i)}{(v_j-w_i)}\l(a(v_j)\prod_{k\neq
i}^M(v_j-w_k+\eta)- d(v_j)\prod^M_{k\neq i}(v_j-w_k-\eta)\r)\label{F}
\eeq 
and $\l\{\vec u\r|$ is the {\it left} vector
defined by 
$$\l\{\vec u\r|=\left<\chi\right|B(u_M)\cdots B(u_1) $$
for any choice of $\{u_i\}$. 
Above,  $\{ w_i \}$ provide a solution to the Bethe
ansatz
equations (\ref{zgbae}) and the parameters $\{ v_j \}$ are arbitrary.
In using the Slavnov formula it is assumed that the pseudovacuum state
has norm equal to one. Defining 
$$G=F.\Gamma$$ 
where $\Gamma$ is a diagonal matrix with entries  
$$\Gamma_{ij}=\delta_{ij}\frac{(v_j-w_j)}{\prod_{k=1}^M(v_j-w_k)} $$ 
the Slavnov formula may be expressed in the equivalent form 
\beq S(\vec w:\,\vec v)=\frac{\prod_{p=1}^M\prod_{q\neq
p}^M(v_p-w_q)}{\prod_{k>l}^M(v_k-v_l)\prod_{i<j}^M(w_i-w_j)}
\det G \label{slavnov2}
\eeq  
with 
\bea &&G_{ij}=\frac{\eta d(w_i)(v_j-w_j)}{(v_j-w_i)^2}\nn \\ 
&&~~~~~~\times \l(a(v_j)\prod_{k\neq
i}^M\frac{(v_j-w_k+\eta)}{(v_j-w_k)}- d(v_j)\prod^M_{k\neq i}
\frac{(v_j-w_k-\eta)}{(v_j-w_k)}\r). \label{G} \eea     
We will find it convenient to use both forms
(\ref{slavnov1},\ref{slavnov2}) of the Slavnov formula. 

The Yang-Baxter algebra $Y[gl(2)]$ admits a conjugation operation 
$\dagger:Y[gl(2)] \rightarrow Y[gl(2)]$   defined by
$$A(u)^\dagger=A(u^*),~~~B(u)^\dagger=C(u^*),~~~C(u)^\dagger=B(u^*),~~~
D(u)^\dagger=D(u^*)$$
and extended to all of $Y[gl(2)]$ through
$$(\theta.\phi)^\dagger=\phi^\dagger.\theta^\dagger, ~~~
\forall\, \theta,\,\phi\,\in Y[gl(2)]$$
such that the defining relations (\ref{rels}) are preserved.
Above, $*$ is used to denote complex conjugation. 
Consequently the right vector 
\bea \l|\vec v\r\}&=&\l\{\vec v\r|^\dagger\nn \\
&=&B(v_1^*)^\dagger\cdots
B(v_M^*)^\dagger\l|\chi\r>\nn \eea   
is also an eigenvector of the transfer matrix whenever the Bethe ansatz
equations for the parameters $\{v_i\}$ are satisfied.
However, it is apparent that the $\Z$-graded realisations
(\ref{zg1},\ref{zg2})
we have introduced are not unitary, and generally
$$ \l<\vec v\r|=\l|\vec v\r>^\dagger 
\neq \l\{\vec v\r|. $$
On the other hand, numerical analysis for the 
models (\ref{jo},\ref{acham},\ref{abcham},\ref{bcs}) 
indicates
that for fixed particle numbers, and
generic values of the coupling parameters, the
energy spectrum is free of degeneracies. This is presumably due to the
fact that the only Lie algebra symmetries for these models are $u(1)$
invariances corresponding to conservation of particle numbers, and the
non-degenerate spectra are examples of Hund's non-crossing rule
\cite{hund,vnw}.
We also assert that for a given $\{v_i\}$, satisfying the
Bethe ansatz equations (\ref{jobae}), this set of parameters is
equivalent to $\{v_i^*\}$;
i.e., $v_i^*=v_j$ for some $j=1,\cdots,M$.
It is clear that for $\{v_i\}$ satisfying (\ref{bae}),
so does $\{v_i^*\}$ (in all our examples, 
as the Hamiltonians are real, the functions $a(u)$ and $d(u)$
are real, as will be seen below). 
Since the eigenvalues of the Hamiltonian are real, we
have
$$E(\vec v)=E^*(\vec v)=E(\vec {v^*}). $$
Under the belief that the spectrum is multiplicity free, we then deduce
$\{v_i\}\sim \{v_i^*\}$.
Whenever this is the case, we can conclude that the eigenvectors are
real and 
\beq \l<\vec v\r| =\zeta(\vec v) \l\{\vec v\r| \label{real}\eeq  
for some non-zero real-valued 
scalar $\zeta(\vec v)$.  Thus the Slavnov formula can be invoked for
the computation of form factors. 
Throughout we will always assume (\ref{real}) to be the case, 
which implies that
\beq \l\{\vec w|\vec v\r>=0 ~~~~~~{\rm for}~~~\vec v\neq \vec w
\label{orthog} \eeq  
whenever $\{v_i\}$ and $\{w_j\}$ both satisfy the Bethe ansatz
equations. This result (\ref{orthog}) 
can be proved directly, independent of (\ref{real}),  
as shown in \cite{ks99}.  
Note when $\vec w=\vec v$ we need to take a limit for the diagonal entries of
$F$ to compute the square of the norm. This yields 
\bea F_{ii}&=&d(v_i)\l(a'(v_i)\prod_{k=1}^M(v_i-v_k+\eta)-d'(v_i)
\prod_{k=1}^M(v_i-v_k-\eta) \r) \nn \\
&&~~+d(v_i)\l(\sum^M_{l\neq i}
\frac{a(v_i)}{v_i-v_l+\eta}\prod_{k=1}^M(v_i-v_k+\eta)\r.\nn \\
&&~~~~~~~~~~~~~~~~\l.-\sum_{l\neq i}^M
\frac{d(v_i)}{v_i-v_l-\eta}\prod_{k=1}^M(v_i-v_k-\eta)\r),\label{diagF} \eea 
where the prime denotes the  derivative. 

From the Slavnov formula the matrix elements of the operators
$B(u),\,C(u)$ follow directly, as the set of parameters $\{v_i\}$ are
arbitrary. We will now derive an expression for the
form factors of the operator $D(u)$, which will prove useful for later
calculations. From the expression (\ref{d}) for the action of $D(u)$   
on an arbitrary Bethe vector we may deduce that for both $\{v_i\}$ and
$\{w_j\}$ satisfying
the Bethe ansatz equations
\bea 
&&\l\{\vec w|D(u)|\vec v\r>
\nn \\
&&~~=  
d(u)\l(\prod_{p=1}^M\frac{u-v_p-\eta}{u-v_p}\r)\l\{\vec w|\vec v\r> 
\nn \\
&&~~~~~~+\sum_{q=1}^M\frac{\eta d(v_q)}{u-v_q}\l(\prod_{j\neq q}^M 
\frac{v_q-v_j-\eta}{v_q-v_j}\r)\l\{\vec w|C(u)|\vec v_i\r> \nn \\ 
&&~~=
d(u)\l.\l(\prod_{p=1}^M\frac{u-w_p-\eta}{u-w_p}\r)\r[S(\vec w:\,\vec v)\nn
\\
&&~~~~~~~+\sum_{q=1}^M\frac{\eta d(v_q)}{(u-v_q)d(u)}\l(\prod_{k=1}^M
\frac{u-w_k}{u-w_k-\eta}\r)\l(\prod_{j\neq q}^M
\frac{v_q-v_j-\eta}{v_q-v_j}\r) \nn \\  
&&~~~\l.\phantom{\l(\prod^M_{i}\r)}\times  
S(\vec w:\,v_1,\cdots,v_{q-1},u,v_{q+1},
\cdots,v_M)\r] \nn \\  
&&~~=\frac{d(u)}{\prod^M_{k>l}(v_k-v_l)\prod^M_{i<j}(w_i-w_j)}
\l(\prod_{p=1}^M\frac{u-w_p-\eta}{u-w_p}\r)
\l(\det F+\Theta(u) \sum_{q=1}^M \det F^{(q)}\r)\nn\eea 
where the matrices $F^{(q)}$ are defined by 
\bea  
F^{(q)}_{pl}&=& F_{pl}~~~~{\rm for~}\,l\neq q 
\nn \\ F^{(q)}_{pq}&=& Q_{pq},  \label{fj} \eea
$Q$ is the rank one matrix with elements
$$Q_{ij}=\frac{\eta d(w_i)d(v_j)}{(u-w_i)(u-w_i-\eta)}
\l(1-\frac{a(u)}{d(u)}\l(\prod_{k\neq
i}^M\frac{u-w_k+\eta}{u-w_k-\eta}\r)
 \r)\prod_{l=1}^M(v_j-v_l-\eta)  $$
and 
\beq    \Theta(u)=\l(\prod_{k=1}^M \frac{u-w_k}{u-v_k}\r) .\label{theta}
\eeq 
Using the fact that for $\vec v\neq \vec w, \,\det F=0$, whereas for 
$\vec v=\vec w$, $\Theta(u)=1$,  allows us to write 
\bea 
&&\l\{\vec w|D(u)|\vec v\r>
\nn \\ 
&&~~=\frac{\Theta(u) d(u)}{\prod^M_{k>l}(v_k-v_l)\prod^M_{i<j}(w_i-w_j)}
\l(\prod_{i=1}^M\frac{u-w_i-\eta}{u-w_i}\r)
\l(\det F+ \sum_{j=1}^M \det F^{(j)}\r)\nn \\  
&&~~=\frac{\Theta(u) d(u)}{\prod^M_{k>l}(v_k-v_l)\prod^M_{i<j}(w_i-w_j)}
\l(\prod_{i=1}^M\frac{u-w_i-\eta}{u-w_i}\r)
\det \l(F+Q\r). \label{med}  \eea 
The last line above follows from the fact that if $X$ is any
$M\times M$ matrix and $Y$ is any rank one $M\times M$ matrix then  
$$\det(X+Y)=\det X+\sum_{j=1}^M \det X^{(j)}$$  
where 
\bea X^{(j)}_{ij}&=& Y_{ij} \nn \\ 
X^{(j)}_{kl}&=& X_{kl}~~~~{\rm for~}\,j\neq l 
. \nn \eea 
A similar result can be derived for the form factors of $A(u)$.
Alternatively, one can obtain them from (\ref{tme},\ref{med}) through  
\beq \l\{\vec w|A(u)|\vec v\r>=\Lambda(u,\vec v)\l\{\vec w|\vec v\r>
-\l\{\vec w|D(u)|\vec v\r>.\label{mea} \eeq  

The Slavnov formula can be extended to include generic $\Z$-graded
realisations as noted in \cite{zlgm}. This is achieved by simply replacing
$a(u)$ and $d(u)$ with $a(u,M)$ and $d(u,M)$ in
(\ref{F},\ref{G}). All the results derived above also extend 
analogously to the $\Z$-graded case.

\subsection{The quasi-classical limit}
Assuming that the quasi-classical limit exists and in particular
\bea a(u)&=&1+\eta\aa (u)+o(\eta^2) \nn \\
d(u)&=&1+\eta\dd (u)+o(\eta^2) \nn \eea 
it is straightforward to obtain the Slavnov formula in the
quasi-classical limit. We obtain directly from (\ref{slavnov1},\ref{slavnov2}) 
\bea \S(\vec w:\,\vec v)
&=&\l<\chi|\B(w_M)\cdots \B(w_1)
\C(v_1)\cdots \C(v_M)|\chi\r> \nn \\
&=&\l<\chi|\B(v_M)\cdots \B(v_1)
\C(w_1)\cdots \C(w_M)|\chi\r> \nn \\
&=&\frac{\prod^M_{p=1}\prod^M_{q\neq p}(v_p-w_q)}{\prod^M_{k>l}(v_k-v_l)
\prod^M_{i<j}(w_i-w_j)}\det \G \nn \\ 
&=&\frac{1}{\prod^M_{k>l}(v_k-v_l)
\prod^M_{i<j}(w_i-w_j)}\det \F \label{qcslavnov} \eea 
where the entries of the $M\times M$ matrices $\G$ and $\F$ are given by 
\bea 
&&\G_{ij}=\left(\aa(v_j)-\dd(v_j)+\sum_{k\neq i}^M
\frac{2}{v_j-w_k}\r)\frac{(v_j-w_j)}{(v_j-w_i)^2} \label{qcg}  \\ 
&&\F_{ij}=\left(\aa(v_j)-\dd(v_j)
+\sum_{k\neq i}^M \frac{2}{v_j-w_k}\r)
\frac{\prod_{l=1}^M(v_j-w_l)}{(v_j-w_i)^2}.  \label{qcf} \eea     
Above, the parameters $\{w_i\}$ are to satisfy the quasi-classical limit
of the Bethe ansatz equations
\beq 
\aa(w_i)-\dd(w_i)=\sum^M_{k\neq i}\frac{2}{w_k-w_i},~~~~~i=1,\cdots,M
\label{qcbae} \eeq  
while the set $\{v_j\}$ are arbitrary. 

Specialising to the case when 
$\{v_i\}=\{w_i\}$ leads to the formula
$$\S(\vec v:\,\vec v)=\det \KK$$ 
where 
\bea 
\KK_{ii}&=&\aa'(v_i)-\dd'(v_i)-\sum^M_{k\neq i}\frac{2}{(v_i-v_k)^2}        
\nn \\
\KK_{ij}&=&\frac{2}{(v_i-v_j)^2}~~~~~~~~~~~~{\rm for~}\,i\neq j. \nn  
\eea 

We can perform a similar treatment to yield the
quasi-classical limit of (\ref{med}). 
The terms in
$\eta^{2M}$ give the scalar product of the states in the quasi-classical
limit. The terms in
$\eta^{2M+1}$ give not only the form factor for $\D(u)$ but also the
next order terms in the expansion of the scalar product, so some care
needs to be taken in order to identify the appropriate terms. The result
is 
\bea \l\{\vec w|\D(u)|\vec v\r>&=&
\frac{\Theta(u)}{\prod^M_{k>l}(v_k-v_l)\prod^M_{i<j}(w_i-w_j)}  
\nn \\
&&~~~\times\l(\l(\dd(u)-\sum_{p=1}^M\frac{1}{u-w_p}\r)\det \F 
+\sum_{q=1}^M\det\F^{(q)}\r)
\label{qcmed} \eea 
where $\F^{(q)}$ is defined in terms of $\F$ and $\Q$ 
in analogy with (\ref{fj}) and 
\beq \Q_{ij}(u)=\frac{\prod_{l\neq j}^M(v_j-v_l)}{(u-w_i)^2}  
\l(\aa(u)-\dd(u)+\sum_{k\neq i}^M\frac{2}{u-w_k}\r). \label{qcq} \eeq 

\section{Exact solution of the models}
\subsection{Solution for the 
model of two Josephson coupled Bose-Einstein condensates}

It is an algebraic exercise to show
that the Hamiltonian (\ref{jo}) is related with the
matrix $\tilde{t}(u,0)=\tilde{A}(u,0)+\tilde{D}(u,0)$ obtained through
(\ref{zg1}) via 
\bea
H&=&
-\frac {\E_J}{2}
\left [\tilde{t} (0, 0) -\eta^{-2} +(\alpha N +\beta)^2
-\eta \sigma N -\eta \delta N^2 \right]
\nn \eea
where we have chosen $\o(N)= \alpha N+\b$ and
the coupling constants are
identified as
\bea
\eta^2&=& \frac{2({U_{11}+U_{22}-U_{12}})}{\E_J}, \no \\
\alpha &=&\frac{U_{11}-U_{22}}{\eta \E_J},\no \\
\beta &=& \frac{\mu_1 -\mu_2}{\eta \E_J}, \no \\
\sigma &=& \frac{\mu_1+\mu_2}{\eta \E_J}, \no \\
\delta&=& \frac{U_{11}+U_{22}}{\eta \E_J}. \label{parameters} \eea
Noting that
$$N=\eta ^{-1} \frac{\rmd\tilde{t}}{\rmd u}(0,0), $$
the above demonstrates that the Hamiltonian (\ref{jo}) is expressible
solely in terms of the matrix $\tilde{t}(u,0)$ and its derivative.

Since $[H,\,N]=0$, the
Hamiltonian is block diagonal in the Fock basis (\ref{fock}). Thus on
a subspace of the Fock space with fixed particle number $N$, the
diagonalisation of $\tilde{t}(u,0)$ is equivalent to the
diagonalisation
of $t(u,0)$ presented earlier in the Bethe ansatz framework. It is
easily determined that for this case, the total particle number $N=M$ and   
\bea &&a(u,N)=u^2-(\alpha N+\beta)^2, \nn \\
&&d(u,N)=\eta^{-2}. \nn \eea 
From (\ref{zgbae},\ref{zgtme}) we 
deduce the solution of (\ref{jo}) for the energy spectrum to be  
\bea  E(\vec v)&=&
-\frac{\E_J}{2}\left[
\eta^{-2}\prod_{i=1}^N\frac{v_i+\eta}{v_i}-\left(\alpha N+\beta
\right)^2 \prod_{i=1}^N\frac{v_i-\eta}{v_i}   \right. \no \\
&&~~~~~~~~~\left.  -\eta^{-2}+(\alpha N + \b)^2
-\eta \sigma N -\eta \delta N^2
\right]\label{jonrg} \eea
where the parameters $\{v_i\}$ are subject to the Bethe ansatz equations
\beq \eta^2\left(v_i^2-\left(\alpha N+\beta
\right)^2\right)=\prod_{j\neq i}^N\frac{v_i-v_j-\eta}{v_i-v_j+\eta}.
\label{jobae} \eeq

\subsection{Solution  for 
the model of homo-atomic-molecular Bose-Einstein condensates}

In terms of a  realisation of the algebra $su(1,1)$ through
\beq
K^+ = \frac {(a^\dagger)^2}{2},
\, K^- = \frac {a^2}{2},
\, K^z= \frac {2N_a+1}{4},
\label{acrep} \eeq
one may establish the relation between the Hamiltonian (\ref{acham}) and the
corresponding matrix $\tilde{t} (u,0)=\tilde{A}(u,0)+
\tilde{D}(u,0)$ arising from the realisation (\ref{zg2}) of the
Yang-Baxter algebra is
\bea
H&=&\sigma + \delta (N/2+1/4)+\gamma (N/2+1/4)^2
+ 2\eta ^{-2}\Omega \tilde{t} (0,0), \nn \eea
with
$$\frac{\rmd\tilde{t}}{\rmd u}(0,0) = 2-\eta (\eta +\alpha) (N/2+1/4)-\eta
\beta.$$
Above we have chosen
\bea \o(K^z+N^c)&=& \alpha (K^z+N^c) +\beta \nn \\
&=&\alpha (N/2+1/4) +\beta \nn \eea
and the following identification has been made for the coupling
constants
\bea
\eta &=& \frac {4U_{aa}+ U_{cc} -2 U_{ac}}
{2\Omega},\no\\
\alpha &=& \frac {U_{cc}- 4U_{aa}}
{2\Omega},\no\\
\beta &=& \frac {2\mu_c -4\mu_a +4U_{aa}- U_{ac}}
{4\Omega}, \no \\
\sigma&=& \frac{U_{aa}-2 \mu_a}{4} , \no \\
\delta &=& \frac{2\mu_c-U_{ac}}{2} , \no \\
\gamma &=& U_{cc}  . \no \eea
We deduce  
\bea &&a(u,M)=\l(u+\eta\kappa\r)\l(1-\eta
u-\eta\l(\alpha(M+\kappa)+\beta\r)\r) \nn \\
&&d(u,M)=u-\eta\kappa \nn \eea 
and by the same argument as in the previous example, 
we conclude that the exact solution for
the energy spectrum of (\ref{acham}) is determined by
(\ref{zgbae},\ref{zgtme}) which reduces to 
\bea
E(\vec v)&=&\sigma +\delta(M+\kappa)+\gamma(M+\kappa)^2 \no \\
&&+ 2\eta ^{-1}\kappa\Omega \left[(1-\eta\left(\alpha (M+\kappa)+\beta)
\right)
\prod_{i=1}^M
\frac{v_i-\eta}{v_i}-\prod_{i=1}^M\frac{v_i+\eta}{v_i}\right],
\label{acnrg} \eea
where the parameters $\{v_i\}$ satisfy the Bethe ansatz equations
\beq
[1-\eta v_i -\eta (\alpha (M+\kappa)+\beta)]
\left(\frac {v_i+\eta \kappa}{v_i-\eta \kappa}\right)=
\prod^M_{j \neq i} \frac {v_i-v_j-\eta}{v_i-v_j+\eta}.
\label {acbae} \eeq
For the representation (\ref{acrep}) of the $su(1,1)$ algebra there are
two
lowest weight vectors; viz.
the Fock vacuum $\left|0\right>$ and the one particle state $a^\dagger
\left|0\right>$. It follows from (\ref{acrep}) 
that the allowed values for $\kappa$ in
(\ref{acnrg},\ref{acbae}) are $\kappa=1/4,\,3/4.$ This demonstrates
that the
solution of the model depends on whether the total particle number
$N=2M+2\kappa-1/2$
is even or odd, the effects of which on the energy spectrum
can be seen through numerical
analysis (cf. \cite{zlm}).

\subsection{Solution for the 
model of hetero-atomic-molecular Bose-Einstein condensates} 

In order to show the solvability of the model (\ref{abcham}), we adopt the
realisation of the $su(1,1)$ algebra given by
\beq
K^+ = a^\dagger b^\dagger,
\, K^- = ab,
\, K^z= \frac {N_a+N_b+1}{2},
\label{abcrep} \eeq
and observe that the operator $\I=N_a-N_b$ commutes with the $su(1,1)$ algebra
in this representation, hence taking a constant value in any
irreducible representation. Due to the symmetry upon interchanging the
labels $a$ and $b$, we can assume without loss of generality that the
eigenvalues of $\I$ are non-negative.
In particular, note then that the lowest weight
states for this realisation are of the form
$$\left|m\right>=\frac{(a^\dagger)^m}{\sqrt{m!}}
\left|0\right>,~~~~~m=0,1,2,\cdots,\infty $$
and $K^z\left|m\right>=(m/2+1/2)\left|m\right>$.
We conclude that the lowest weight labels $\kappa$ can be taken from
the set $\{1/2,\,1,\,3/2,....\}$ and the eigenvalue of $\I$ in
the irreducible representation labelled by $\kappa$ is $2\kappa-1$.

For this case the relation between the Hamiltonian (\ref{abcham}) and the
corresponding matrix $\tilde{t} (u,0)$ from (\ref{zg2}) is
\bea
H&=&
\sigma+\delta(N/2+1/2)+\lambda(N/2+1/2)^2 \no \\
&&~~+ \rho \I + \nu \I^2 +
\xi \I (N/2+1/2)
+ \eta ^{-2}\Omega \tilde{t} (0,0)
\eea
with
$$\frac{\rmd\tilde{t}}{\rmd u}(0,0) = 2-\eta (\eta +\alpha) (N/2+1/2)-\eta 
\gamma \I-\eta \beta.$$
Above we have chosen
\bea \omega (K^z+N^c)&=& \alpha (K^z+N^c) +\gamma (2\kappa-1)+\beta\nn \\
&=&\alpha (N/2+1/2) +\gamma \I +\beta\nn \eea
and
the coupling constants are related through 
\bea
\eta &=& \frac{U_{aa}+U_{bb}+U_{cc}+U_{ab}-U_{ac}-U_{bc}}{\Omega}  , \nn
\\
\alpha &=&\frac{U_{cc}-U_{aa}-U_{bb}-U_{ab}}{\Omega}  , \nn \\
\beta
&=&\frac{2U_{aa}+2U_{bb}+2U_{ab}-U_{ac}-U_{bc}+2\mu_c-2\mu_a-2\mu_b}
{2\Omega}  , \nn \\
\gamma &=& \frac{2U_{bb}-2U_{aa}+U_{ac}-U_{bc}}{2\Omega}, \no \\
\sigma &=&\frac{U_{aa}+U_{bb}+U_{ab}-2\mu_a-2\mu_b}{4} , \nn \\
\delta &=&\frac{2\mu_c-U_{ac}-U_{bc}}{2}   , \nn \\
\lambda &=&U_{cc}  , \nn \\
\rho &=&\frac{U_{bb}-U_{aa}+\mu_a-\mu_b}{2}  , \nn \\
\nu &=&\frac{U_{aa}+U_{bb}-U_{ab}}{4}  , \nn \\
\xi &=&\frac{U_{ac}-U_{bc}}{2}
.\no
\eea
We find 
\bea &&a(u,M)=\l(u+\eta\kappa\r)\l(1-\eta
u-\eta\l(\alpha(M+\kappa)+\gamma(2\kappa-1)+\beta\r)\r), \nn \\
&&d(u,M)=(u-\eta\kappa) 
\nn \eea 
and the exact solution in this instance reads
\bea
E(\vec v)&=&
\sigma +\delta(M+\kappa)+\lambda(M+\kappa)^2\no \\
&&~~+\rho(2\kappa-1)+
\nu(2\kappa-1)^2+\xi(2\kappa-1)(M+\kappa) \no \\
&& ~~+ \eta ^{-1}\kappa\Omega
\left[(1-\eta(\alpha (M+\kappa)+\gamma(2\kappa-1)+\beta))
\prod_{i=1}^M
\frac{v_i-\eta}{v_i}\right. \no \\
&&~~~~~~~~~~~~~~~~~~~~~~~~~~~~~~~
~~\left.-\prod_{i=1}^M\frac{v_i+\eta}{v_i}\right],
\nn \eea
where the parameters $\{v_i\}$ satisfy the Bethe ansatz equations
\beq
[1-\eta v_i -\eta \left(\alpha
(M+\kappa)+\gamma(2\kappa-1)+\beta\right)]
\left( \frac {v_i+\eta \kappa}{v_i-\eta \kappa}\right)=
\prod^M_{j \neq i} \frac {v_i-v_j-\eta}{v_i-v_j+\eta}.  
\label {abcbae} \eeq
The total atom number is given by $N=2M+2\kappa -1$.

\subsection{Solution for the reduced BCS model}

\def\beq{\begin{equation}}
\def\eeq{\end{equation}}
\def\bea{\begin{eqnarray}}
\def\eea{\end{eqnarray}}
\def\ve{\epsilon}
\def\si{\sigma}
\def\th{\theta}
\def\d{\delta}
\def\l{\left}
\def\r{\right}
\def\b{\beta}


\def\b{\beta}
\def\d{\dagger}
\def\e{\epsilon}
\def\k{\kappa}
\def\l{\lambda}
\def\o{\omega}
\def\t{\tilde{\tau}}
\def\s{S}
\def\T{{\cal T}}

\def\beq{\begin{equation}}
\def\eeq{\end{equation}}
\def\bea{\begin{eqnarray}}
\def\eea{\end{eqnarray}}
\def\ba{\begin{array}}
\def\ea{\end{array}}
\def\no{\nonumber}
\def\le{\langle}
\def\re{\rangle}
\def\lt{\left}
\def\rt{\right}
\def\oR{R^*} \def\upa{\uparrow}
\def\doa{\downarrow}
\def\nn{\nonumber} \def\dag{\dagger}
\def\e{\epsilon}
\def\si{\sigma}
\def\th{\theta}
\def\de{\delta}
\def\l{\left}
\def\r{\right}
\def\b{\beta}
\def\La{\Lambda}
\def\rr{\mathcal{R}}
\def\T{\mathcal{T}}

As an alternative to the BCS mean-field approach, one can appeal to the 
exact solution of the Hamiltonian (\ref{bcs}) as given in 
\cite{richardson,rs}.
Much  later it was shown by Cambiaggio, Rivas and Saraceno \cite{crs} that 
(\ref{bcs})
is integrable in the sense that there exists a set of mutually commutative 
operators which commute with the Hamiltonian. 
Our aim here is to show that both of these features are consequences of
the fact that the Hamiltonian (\ref{bcs}) can be derived using the
quantum inverse scattering method. This result, which was established
in \cite{zlmg,vp}, will be proved  below. 
Before doing so, let us remark that there have been several works on
this problem, including generalisations \cite{ado1,ado2,des}. 
In some cases these
models can be obtained using 
trigonometric/hyperbolic 
versions of the Yang-Baxter algebra. While we will not go
into details here, this generalisation from the procedure described
below is straightforward and is simply a matter of using the
trigonometric/hyperbolic analogue of (\ref{rm}) from the outset.  

We use 
a $c$-number realisation $\g$ of the $L$-operator, defined by 
$\g=\exp(-\alpha\eta {\sigma})$ with
$\sigma={\rm diag}(1,\,-1)$, as well as 
(\ref{ls}) 
to  construct the transfer matrix 
\beq t(u)={\rm tr}_0\left(\g_0L^S_{0{\L}}(u-\e_{\L})\cdots 
L^S_{01}(u-\e_1)\right) \label{bcstm} \eeq  
which is an element of the $\L$-fold tensor algebra of $su(2)$.
Here, 
${\rm tr}_0$ denotes the trace
taken over the auxiliary space, which for convenience is labelled by 0,
while the tensor components of the physical space are labelled
$1,\cdots,\L$. 
Defining 
$$T_j=\lim_{u\rightarrow \e_j}\frac{u-\e_j}{\eta^2} t(u) $$ 
for $j=1,2,...,{\L}$,
we may write in the quasi-classical limit  
$T_j=\tau_j+o(\eta) $ 
and it follows from the commutativity of the transfer matrices that 
$ [\tau_j,\,\tau_k]=0, ~ \forall\, j,\,k. $ 
Explicitly, these operators read 
\beq 
\tau_j=-2\alpha S^z_j+\sum_{k\neq j}^{\L}\frac{\theta_{jk}}{\e_j-\e_k}
\label{cons} \eeq 
with 
$\theta=S^+\otimes S^-+S^-\otimes S^++2S^z\otimes S^z.$ 
The set of operators (\ref{cons}), 
realised in terms of canonical fermion operators, 
are those obtained by Cambiaggio et al.
\cite{crs} to establish the integrability of the reduced BCS model. They
first appeared in the work of Sklyanin \cite{sklyanin1} in
a general context, and are the Gaudin Hamiltonians
\cite{gaudin1,gaudin2} in the presence of a non-uniform magnetic field. 

Next define a Hamiltonian through
\bea  
H&=&-\frac{1}{\alpha}\sum_{j=1}^{\L}\e_j\tau_j
+\frac{1}{4\alpha^3}\sum_{j,k=1}^{\L}\tau_j\tau_k 
+\frac{1}{2\alpha^2}\sum_{j=1}^{\L} \tau_j- \frac{1}{2\alpha}
\sum_{j=1}^{\L}C_j \label{ham} \\ 
&=& \sum_{j=1}^{\L}2\e_jS_j^z-\frac{1}{\alpha}\sum_{j,k=1}^{\L}S_j^+S_k^-
\label{ham1} \eea       
where 
$$C=S^+S^-+S^-S^++2(S^z)^2$$ 
is the Casimir invariant for the $su(2)$ algebra. 
The Hamiltonian is universally integrable since it is clear that 
$[H,\,\tau_j]=0,~~\forall j $, irrespective of the realisations of the 
$su(2)$ algebra in the tensor algebra.  

In order to reproduce the Hamiltonian (\ref{bcs}) we realise 
the $su(2)$ generators through the (spin 1/2) hard-core boson  
 representation (\ref{hcb}); viz
\beq S_j^-=b_j, ~~~
S_j^+=b_j^{\dagger}, ~~~ 
S^z_j=\frac{1}{2}\left(n_{j}-I\right) \label{psr}. \eeq  
In this instance one obtains (\ref{bcs}) (up to the constant term 
$-\sum_{j=1}^\L\e_j$) with $g=1/\alpha$.   
Incorporating 
higher spin  representations of the $su(2)$ algebra yields models
which may be interpreted as coupled BCS systems  
\cite{ado1,ado2,lh,lzmg}. Generally, we can define a representation
of $su(2)$ through  
\beq S_j^-=-\sum_{\sigma\in \Omega}a_{j\sigma}a_{j\overline{\sigma}}, ~~~
S_j^+=\sum_{\sigma\in\Omega}a_{j{\sigma}}^{\dagger}
a_{j\overline{\sigma}}^{\dagger}, 
~~~
S^z_j=\frac{1}{2}\sum_{\sigma\in\Omega}\left(n_{j\sigma}+n_{j\overline{\sigma}}
\pm I\right) \label{idr} \eeq
 where the operators $a_{j\sigma},\,a_{j\sigma}^\dagger$ 
 may be either bosonic (+ sign in $S^z_j$) or fermionic (- sign in
 $S^z_j$). Above, $\sigma\in\Omega$ is 
 a degeneracy label and $\overline{\sigma}\notin\Omega$
 refers to the time-reversed state (i.e., the total degeneracy is twice
 the cardinality of $\Omega$). 
In this instance one recovers the pairing models discussed in \cite{des,ds,dp}.
(For the bosonic case it is convenient to replace $L^S(u)$ with $L^K(u)$ since
(\ref{idr}) is not unitary for bosons.) 
Because of these different possibilities 
we will derive the eigenvalues of the Hamiltonian
(\ref{ham1}) in a general setting. 

For each index $k$ of the tensor algebra in which the transfer matrix acts, 
and accordingly in (\ref{ham1}),
suppose that we represent the $su(2)$ 
algebra through the irreducible representation with lowest weight
(or spin) $-s_k$. Note that we impose no restriction on the allowed
values of $s_k$ in order to accommodate infinite-dimensional  
representations such as the bosonic case of (\ref{idr}). 
Choosing the pseudovacuum state to be the tensor product of lowest
weight states gives 
\bea a(u)&=&\exp(-\alpha\eta)\prod_{k=1}^\L\frac{u-\e_k-\eta s_k}
{u-\e_k} \nn \\
d(u)&=&\exp(\alpha\eta)\prod_{k=1}^\L\frac{u-\e_k+\eta s_k}{u-\e_k} \nn \eea 
and the eigenvalues of the transfer matrix  
(\ref{bcstm}) as 
\bea
\Lambda(u)&=&\exp(-\alpha\eta)\prod_{k=1}^{\L}\frac{u-\e_k-\eta s_k}{u-\e_k}
\prod_{j=1}^M\frac{u-v_j+\eta}{u-v_j} \nn \\
&&~
+\exp(\alpha\eta)\prod_{k=1}^{\L}\frac{u-\e_k+\eta s_k}{u-\e_k}
\prod_{j=1}^M\frac{u-v_j-\eta}{u-v_j}. \nn  
 \eea
The corresponding Bethe ansatz equations read 
$$\exp(-2\alpha\eta)\prod_{k=1}^{\L}\frac{v_i-\e_k-\eta s_k}
{v_i-\e_k+\eta s_k}
=\prod_{j\neq i}^M\frac{v_i-v_j-\eta}{v_i-v_j+\eta} . $$

The eigenvalues of the conserved operators (\ref{cons}) are obtained
through the appropriate 
terms in the expansion of the transfer matrix eigenvalues  
in the parameter $\eta$. This yields the following result 
for the eigenvalues $\lambda_j$ of $\tau_j$ 
\beq \lambda_j= \left(2\alpha +\sum_{k\neq j}^{\L}\frac{2s_k}{\e_j-\e_k} 
-\sum_{i=1}^M 
\frac{2}{\e_j-v_i}\right)s_j \label{eig} \eeq  
such  that the parameters $\{v_j\}$ satisfy the quasi-classical limit of 
the Bethe ansatz equations 
\beq 
2\alpha+ \sum_{k=1}^{\L}\frac{2s_k}{v_i-\e_k}
=\sum_{j\neq i}^M \frac{2}{v_i-v_j}.
\label{bcsbae} \eeq  
For $s_k=1/2,\,\,\forall\,k$ these equations were found in
\cite{richardson} through a different technique. 

Through (\ref{eig}) we can now determine the energy eigenvalues of 
(\ref{ham1}). It is useful to note the following identities 
\bea 
&&2\alpha \sum_{j=1}^{M}v_j+2\sum_{j=1}^M\sum_{k=1}^{\L}\frac{v_js_k}{v_j-\e_k}
=M(M-1) \nn \\
&&\alpha  M+\sum_{j=1}^M\sum_{k=1}^{\L}\frac{s_k}{v_j-\e_k}=0 \nn \\
&&\sum_{j=1}^M\sum_{k=1}^{\L}\frac{v_js_k}{v_j-\e_k} -
\sum_{j=1}^M\sum_{k=1}^{\L}\frac{s_k\e_k}{v_j-\e_k}
=M\sum_{k=1}^{\L}s_k. 
\nn\eea 
Employing the above it is deduced that 
\bea 
\sum_{j=1}^{\L}\lambda_j &=&2\alpha\sum_{j=1}^{\L}s_j-2\alpha M \nn \\
\sum_{j=1}^{\L}\e_j\lambda_j&=&2\alpha \sum_{j=1}^{\L}\e_j s_j
+\sum_{j=1}^{\L}\sum_{k\neq j}^{\L}  
s_js_k-2M\sum_{k=1}^{\L}s_k-2\alpha\sum_{j=1}^M v_j +M(M-1) \nn \eea 
which, combined with the eigenvalues $2s_j(s_j+1)$ for the Casimir invariants
$C_j$, yields from (\ref{ham},\ref{eig}) the energy eigenvalues 
\beq E=2\sum_{j=1}^{M} v_j -2\sum_{k=1}^{\L}s_k \e_k. \label{bcsnrg} \eeq  
{}From the above expression we see that the quasi-particle excitation
energies are given by twice the Bethe ansatz roots $\{v_j\}$ 
of (\ref{bcsbae}). 
Finally, let us remark that the eigenstates 
obtained in taking the quasi-classical limit assume  the form 
$$\left|\vec v\right>=\prod_{i=1}^M\l(\sum_{j=1}^\L\frac{S^+_j}{v_i-\e_j}
\r)\left|0\right> $$
where $\{v_i\}$ satisfy (\ref{bcsbae}), 
which are the same as those obtained by Richardson \cite{richardson} in
the case of the reduced BCS model. 

An alternative approach to the exact solution of the reduced BCS model, 
which produces both the
eigenstates and the eigenvalues of the conserved operators, was given by Sierra
\cite{sierra} using conformal field theory given by the $SU(2)_k$-WZW
model in the limit when the level $k$ approaches $-2$. There also exists
an intriguing analogy for the reduced BCS model 
from two-dimensional electrostatics \cite{dep,admor}.  

\section{Exact calculation of form factors} 

Through use of the Slavnov formula (\ref{slavnov1},\ref{slavnov2}) 
and (\ref{med},\ref{mea}) 
we have 
explicit determinant representations for the form factors of
$A(u),\,B(u), \,C(u)$ and $D(u)$. 
Given any operator, we would like to be able to
express it solely in
terms of these operators, which we call the inverse problem. 
Solution of the inverse problem then 
permits us to determine the form factors for
that operator. In the following we will show in several examples how
this can be achieved. In some cases we will
restrict our analysis to some subclass of the
models (\ref{jo},\ref{acham},\ref{abcham}).

\subsection{Form factors for the 
model of  two Josephson coupled 
Bose-Einstein condensates}\label{fff}

Specialising the Slavnov formula
(\ref{slavnov1},\ref{slavnov2}) 
to the case of the Hamiltonian (\ref{jo}) 
gives the matrix elements of $F$ and $G$ as
$$ 
F_{ij}=\frac{\eta^{-1}}{(v_j-w_i)}
\l((v_j^2-(\alpha N+\beta)^2))\prod_{k\neq i}^N
\frac{(v_j-w_k+\eta)}{(v_j-w_k)}- \eta^{-2}
\prod^N_{k\neq i} \frac{(v_j-w_k-\eta)}{(v_j-w_k)}\r),$$ 
$$G_{ij}=\frac{\eta^{-1}(v_j-w_j)}{(v_j-w_i)^2}
\l((v_j^2-(\alpha N+\beta)^2))\prod_{k\neq i}^N 
\frac{(v_j-w_k+\eta)}{(v_j-w_k)}- \eta^{-2}
\prod^N_{k\neq i} \frac{(v_j-w_k-\eta)}{(v_j-w_k)}\r). $$  
In order to apply the Slavnov formula for the computation of 
wavefunction norms, we need to determine the functions $\zeta(\vec v)$,
introduced in (\ref{real}). 
We can write
\bea \l|\vec v\r>&=&C(v_1,1)\cdots C(v_N,N)\l|0\r> \nn \\  
&=&\sum_{k=0}^N x_k\l(a_1^\dagger\r)^k\l(a_2^\dagger\r)^{N-k}\l|0\r> \nn
\eea  
for some scalar functions $x_i$. 
We deduce from the explicit form of $C(u,j)$ that 
\bea x_0&=&\eta^{-N} \nn \\
x_N&=&\prod_{i=1}^N\l(v_i-\alpha N-\beta\r).  \nn \eea 
On the other hand we have 
\bea \l|\vec v\r\}&=&B(v_1,1)^\dagger\cdots B(v_N,N)^\dagger\l|0\r> \nn
\\
&=&\sum_{k=0}^N y_k \l(a_1^\dagger\r)^k\l(a_2^\dagger\r)^{N-k}\l|0\r>
\nn
\eea  
with 
\bea y_0&=&\prod_{i=1}^N\l(v_i+\alpha N+\beta\r) \nn \\
y_N&=&\eta^{-N}. \nn \eea  
Using the identity (\ref{identity}) applied to the present case 
$$\prod_{i=1}^N\l(v_i^2-\l(\alpha N+\beta\r)^2\r)=\eta^{-2N} $$ 
gives 
\bea x_0&=&\eta^N\prod_{i=1}^N\l(v_i-\alpha N-\beta\r)y_0 \nn \\
x_N&=&\eta^N\prod_{i=1}^N\l(v_i-\alpha N-\beta\r)y_N \nn \eea  
which shows that 
$$\zeta(\vec v)= \eta^N\prod_{i=1}^N\l(v_i-\alpha N-\beta\r). $$  
The square of the wavefunction norms are then given by 
\bea ||\vec v||&=&\l<\vec v|\vec v\r> \nn \\
&=&\zeta(\vec v) S(\vec v:\,\vec v) \nn \eea 
where $S(\vec v:\vec v)$ is expressible in terms of $F$ or $G$.

{}From (\ref{zg1}) we can see that
solution to the inverse problem is achieved through  
\bea a_1^\dagger&=&\lim_{u\rightarrow \infty}\frac{1}{u}C(u), \nn \\
a_2&=&\lim_{u\rightarrow\infty}\frac{1}{u}B(u). \nn \eea 
Using the Slavnov formula we  have, 
for $\l|\vec v\r>,\,\l|\vec w\r>$ both eigenstates of the Hamiltonian, 
\bea \l<\vec v\r|a_1\l|\vec w\r>&=&\l<\vec w\r|a^\dagger_1\l|\vec v\r>\nn \\
&=&\zeta(\vec w)\l\{\vec w\r|a^\dagger_1\l|\vec v\r> \nn \\
&=&\zeta(\vec w)\lim_{u\rightarrow\infty}\frac{1}{u}
\l\{\vec w\r|C(u)\l|\vec v\r> \nn \\
&=&\zeta(\vec w)\lim_{u\rightarrow \infty}\frac{1}{u}
S(\vec w:\, v_1,\cdots,v_{M-1},u) \nn \\
&=&\frac{\zeta(\vec w)\prod_{p=1}^{M-1}\prod_{q\neq p}^M(v_p-w_q)}
{\prod_{k>l}^{M-1}(v_k-v_l)
\prod_{i<j}^M(w_i-w_j)}\det \overline{G}    
\nn \eea 
where 
\bea \overline{G}_{ij}&=& G_{ij}~~~~~~~~{\rm for}~~~~j\neq M, \nn \\
\overline{G}_{iM}&=&\eta^{-1}. \nn \eea 
In a similar way we find 
\bea \l<\vec w\r|a_2^\dagger\l|\vec v\r>&=&\l<\vec v\r|a_2\l|\vec w\r>\nn \\
&=&\zeta(\vec v)\l\{\vec v\r|a_2\l|\vec w\r> \nn \\
&=&\zeta(\vec v)\lim_{u\rightarrow\infty}\frac{1}{u}
\l\{\vec v\r|B(u)\l|\vec w\r> \nn \\
&=&\zeta(\vec v)\lim_{u\rightarrow\infty}\frac{1}{u}
\l\{\vec w\r|C(u)\l|\vec v\r> \nn \\
&=&\zeta(\vec v)\lim_{u\rightarrow \infty}\frac{1}{u}
S(\vec w:\,v_1,\cdots,v_{M-1},u) \nn \\
&=&\frac{\zeta(\vec v)\prod_{p=1}^{M-1}\prod_{q\neq p}^M(v_p-w_q)}
{\prod_{k>l}^{M-1}(v_k-v_l)
\prod_{i<j}^M(w_i-w_j)}\det \overline{G}   
\nn \eea 
with $\overline{G}$ as above. 

In Josephson's
original proposal \cite{josephson1,josephson2} for tunneling of Cooper
pairs through an insulating barrier, the effect is a manifestation of
the relative phase difference of the wavefunctions for the two
superconductors. Josephson exploited the fact that the BCS variational
wavefunction (\ref{bcswf}) is not an eigenstate of the total particle number,
and as phase and particle number are canonically conjugate variables, a
well defined relative phase could be assigned. For the model (\ref{jo})
there are technical difficulties which prevent a simple definition for
the phase variable \cite{leggett,phase}. Consequently, the 
expectation values for the Josephson tunneling current
$$\j=i(a_1^\d a_2-a_2^\d a_1) $$
as well as $\n$ and $\n^2$, where
$\n=N_1-N_2$ is the relative particle number
operator, are of primary interest. 
In principle, these can all be expressed in terms of the form
factors for $a_1,\,a^\dagger_1,\,a_2$ and $a^\dagger_2$ through
completeness relations. This would yield expressions comprised of sums
of determinants. 
However, in the case when
\beq U_{11}=U_{22},\, ~~~~~~~~~\mu_1=\mu_2, \label{constraint} \eeq
which results in $\alpha=\beta=0$ from (\ref{parameters}),
we can use a direct method to yield the form factors for $\j,\,\n$ and 
$\n^2$, expressed as single determinants \cite{lz}.
The reason we can achieve this under the constraint (\ref{constraint}) 
is that in this case the Hamiltonian acquires the additional symmetry 
$$[P,\,H]=0$$
where $P$ is the permutation operator
defined by the action on the Fock basis
$$P.(a^{\d}_1)^j(a^{\d}_2)^k\left|0\right>
=(a^{\d}_1)^k(a^{\d}_2)^j\left|0\right>. $$ 
This means that the energy eigenstates 
are also eigenstates of $P$,  
and moreover,
$P^2=I$ shows that $P$ has eigenvalues $\pm 1$. 
Only by exploiting this symmetry do the form factors for $\n,\,\n^2$ and
$\j$ become accessible. 

As mentioned earlier, the realisation of $Y[gl(2)]$ used to derive
the model (\ref{jo}) is not unitary. It is however equivalent to a
unitary representation when (\ref{constraint}) is satisfied 
in the sense that 
$$C^\d (u)=PB(u^*)P. $$ 
Consider for $\{v_i\}$ satisfying the Bethe ansatz equations  
\bea \left<\vec v|\vec v\right>&=&\zeta(\vec v)S(\vec v:\,\vec v)\nn \\
&=&\zeta(\vec v)\left<0|B(v_N)\cdots B(v_1)C(v_1)
\cdots C(v_N)|0\right> \nn \\  
&=&\zeta(\vec v)\left<0|PC^{\d}(v_N^*)P...PC^{\d}(v_1^*)P
C(v_1)...C(v_N)|0\right> \nn \\
&=&\zeta(\vec v)\left<0|C^{\d}(v_N^*)...C^{\d}(v_1^*)P
C(v_1)...C(v_N)|0\right> \nn \\  
&=&\zeta(\vec v)\left<\vec v\r|P\l|\vec v\right> \nn \eea 
which shows that  $\zeta(\vec v)=\pm 1$ is the eigenvalue of $P$; viz. 
$$P\left|\vec v\right>=\zeta(\vec v)\left|\vec v\right>. $$ 
From the Slavnov formula, the squares of the norms of the eigenstates 
in this limit 
\bea ||\vec v||^2&=&\left<\vec v|\vec v\right> \nn \\
&=&\left|S(\vec v:\vec v)\right| \nn \eea   
are obtained directly.

We define
$$\Xi=A(0)-D(0)=\eta^2N_1N_2 +i\j -\eta^{-2}. $$
Letting $\l|\vec v\r>$ and $\l|\vec w\r>$ be eigenstates of the
Hamiltonian we can appeal to (\ref{med}) and (\ref{mea}) to find 
\beq 
\left<\vec w|\Xi |\vec v\right> 
=\frac{-\zeta(\vec v)\zeta(\vec w)\eta^{N-2}\prod^N_{i=1}(w_i+\eta)}
{\prod^N_{k>l}(v_k-v_l)\prod^N_{i<j}(w_i-w_j)}
\det \left(F+2Q \right) 
  \label{ff}  \eeq 
where the elements of $Q$ read  
$$Q_{ij}=\frac{\eta^{-3}\prod^N_{l=1}(v_j-v_l-\eta)}{w_i(w_i+\eta)}. $$ 
We remark that because the basis states are
also Hamiltonian eigenstates, it is straightforward to write down the
time-dependent form factors 
\beq \left<\vec w|\Xi(t)|\vec v\right>=\exp(-it\left(E(\vec w)-E(\vec v)
\right))\left<\vec w|\Xi|\vec v\right> \nn  \eeq  
where the energies are given by (\ref{jonrg}), with $\alpha=\beta=0$. 

Remarkably, from equation (\ref{ff}) all the 
form factors for
$\n,\,\n^2$ and $\j$ can be obtained. This is achieved by exploiting the
symmetry of the Hamiltonian under $P$.
We begin with the following result, which is easily proved. 
If $\zeta(\vec w)\neq \zeta(\vec v)$
then  
$$\left<\vec w|N_1N_2|\vec v\right>=0. $$ 
If $\zeta(\vec w)=\zeta(\vec v)$ then 
$$\left<\vec w|\j|\vec v\right>=0. $$ 
The result follows from the observation
\bea PN_1N_2&=&N_1N_2P, \nn \\
P\j&=&-\j P. \nn \eea 

We now find that 
$$\left<\vec w|N_1N_2|\vec v\right> = 
\eta^{-2}\left<\vec w|\Xi|\vec v\right>
+\eta^{-4}\left<\vec w|\vec v\right> $$ 
if $\zeta(\vec w)=\zeta(\vec v)$, and is zero otherwise. Also  
$$\left<\vec w|\j|\vec v\right> = -i\left<\vec w|\Xi|\vec v\right>
$$
if $\zeta(\vec w)\neq \zeta(\vec v)$, and is zero otherwise. 

The above shows that the form factors for $\j$ are obtained directly from those 
of $\Xi$. Those for $\n^2$ also follow, since we have 
$\n^2={N}^2-4N_1N_2$ and the Hamiltonian eigenstates are also   
eigenstates
of the number operator  $N$. Thus 
\bea \left<\vec w|\n^2|\vec v\right>&=&N^2\left<\vec w|\vec v\right> 
-4\left<\vec w|N_1N_2|\vec v\right>\nn \\
&=&
(N^2-4\eta^{-4})\l<\vec w|\vec v\r>-4\eta^{-2}\l<\vec w|\Xi|\vec v\r> \nn
\eea 
if $\zeta(\vec w)=\zeta(\vec v)$, and zero otherwise. 
To obtain the form factors for $\n$, we use the fact that 
$\j$ is the time derivative of $\n$, so  
$$\j=\frac{i}{\E_J}  [\n,\,H]$$ 
which gives 
\bea \left<\vec w|\n|\vec v\right> &=&\frac{i\E_J}
{E(\vec w)-E(\vec v)} \left<\vec w|\j|\vec v\right>\nn \\
&=&\frac{\E_J}{E(\vec w)-E(\vec v)}\l<\vec w|\Xi|\vec v\r> \nn \eea 
if $\zeta(\vec w)\neq\zeta(\vec v)$ and zero otherwise.  

The expectation values  
$$\left<\theta\right>_\Psi=\frac{\left<\Psi|\theta|\Psi\right>}
{\left<\Psi|\Psi\right>}$$ where $\theta=\n,\,\n^2$ or $\j$, and 
$\left|\Psi\right>$ is an  arbitrary state, can be expressed
in terms of the form factors through completeness relations, in a time
dependent fashion. 
In particular, for a given $\left|\Psi\right>$ 
the quantum fluctuations of the relative number operator
$$\Delta(\Psi;\n)=\left<\n^2\right>_\Psi
-\left<\n\right>_\Psi^2 $$
can be computed from these results. 

The extension of these results to the general case 
without the imposition of the constraint
(\ref{constraint}) remains an open problem.  

\subsection{Form factors for the models of 
atomic-molecular Bose-Einstein condensates} 

As both models for atomic-molecular Bose-Einstein condensates are
derived from the same $L$-operator (\ref{zg2}), we may treat the two
models
simultaneously.
In analogy with the previous model, we can deduce  
\bea 
&&\zeta(\vec v)=\prod_{i=1}^M\l(\frac{\eta\kappa+v_i}{\eta
\kappa -v_i}\r) \nn \\ 
&&c^\dagger = \eta^{-1}\lim_{u\rightarrow \infty}\frac{1}{u}C(u) \nn \\
&&K^+=-\eta^{-2}\lim_{u\rightarrow\infty}\frac{1}{u}\l(B(u)^\dagger + 
C(u) \r). \nn
\eea 
This leads to the following form factors when $\{w_i\}$ and $\{v_j\}$ 
both satisfy the Bethe ansatz equations 
\bea 
\l<\vec v,\kappa |c|\vec w,\kappa \r>
&=&\l<\vec w,\kappa |c^\dagger|\vec v,\,\kappa\r> \nn \\
&=&\eta^{-1}\zeta(\vec w)\lim_{u\rightarrow\infty}\frac{1}{u}
S(\vec w:\,v_1,\cdots,v_{M-1},u), \nn \\
\l<\vec v,\kappa |K^-|\vec w,\kappa \r>
&=&\l<\vec w,\kappa |K^+|\vec v,\kappa\r> \nn \\
&=&-\eta^{-2}\l(\zeta(\vec w)+\zeta(\vec v)\r)
\lim_{u\rightarrow\infty}\frac{1}{u}    
S(\vec w:\,v_1,\cdots,v_{M-1},u). \label{ffk} \eea 
Note that for this class of models we include the label $\kappa$ in
the Bethe states in
order to identify the pseudovacuum used for the Bethe ansatz
calculation.

Realising the $su(1,1)$ algebra in terms of the Heisenberg algebra as in
(\ref{acrep}) or (\ref{abcrep}) gives form factors for the models
(\ref{acham}) and
(\ref{abcham}) respectively.
Certain form factors for the single particle atomic
creation and annihilation operators can also be obtained by
using the fact that for these models there are multiple possible
pseudovacuum states for the Bethe ansatz calculations.
For the model (\ref{acham}) we have
\bea
\left<\vec v,1/4|a|\vec w,3/4\right>
&=&
\left<\vec w,3/4|a^\dagger|\vec v,1/4\right> \nn \\
&=& \left <\vec w,3/4|\vec v,3/4\right>,
\nn \\
\left<\vec v,3/4|a|\vec w,1/4\right>
&=&\left<\vec w,1/4|a^\dagger|\vec v,3/4\right> \nn \\
&=& \left <\vec w,1/4|a^\dagger a^\dagger|\vec v,1/4\right>
\nn \\
&=&
2\left <\vec w,1/4|K^+|\vec v,1/4\right>.
\label{naff1} \eea
In the case of the model (\ref{abcham}) we find
\bea
\left<\vec v,(\kappa-1/2)|a|\vec w,\kappa\right>
&=&\left<\vec w,\kappa|a^\dagger|\vec v,(\kappa-1/2)\right> \nn \\
&=& \sqrt{2\kappa-1}
\left <\vec w,\kappa|\vec v,\kappa \right>, \nn \\
\left<\vec v,(\kappa+1/2)|b|\vec w,\kappa\right>
&=&\left<\vec w,\kappa|b^\dagger| \vec v,(\kappa+1/2)\right>
\nn \\
&=&\frac{1}{\sqrt{2\kappa}}\left <\vec w,\kappa|a^\dagger
b^\dagger|\vec v,\kappa\right>
\nn \\
&=&\frac{1}{\sqrt{2\kappa}}
\left <\vec w,\kappa|K^+|\vec v,\kappa\right>.
\label{naff2} \eea

Note that in the case of (\ref{naff1}), if $\l|\vec v,3/4\r>$ is an
eigenvector of (\ref{acham}), there is no reason to assume that 
$\l|\vec v,1/4\r>$ is also an eigenvector. Hence the formula (\ref{ffk})
cannot be used to evaluate (\ref{naff1}), since in (\ref{ffk}) it is
required that $\l|\vec v,1/4\r>$ is an eigenvector. This is because we can only
establish that (\ref{real}) holds for eigenvectors. A similar situation
applies to (\ref{naff2}). 

In the quasi-classical limit the procedure for computing the form
acquires a simplified form, in which (\ref{naff1},\ref{naff2}) can be
evaluated.  Moreover, the form factors for $K^z$ 
can be obtained, which are seemingly intractable in the general case. 
Below is a detailed account. We set the coupling parameters
$U_{ij}$ to zero in the Hamiltonians (\ref{acham},\ref{abcham}). This
corresponds to the ideal gas limit in the sense that the terms with
coupling $U_{ij}$ describe the $S$-wave scatterings between the
particles. 
Mathematically, this means that $\eta=0$, corresponding
to the quasi-classical limit, and $\omega(x)=\beta$ is constant. We scale
the generating elements $A(u),\,B(u),\,C(u),\,D(u)$ of (\ref{zg2}) 
by a factor of
$1/u$, and in taking the quasi-classical limit
we obtain  
the following realisation of the Gaudin algebra 
\bea 
\A(u)&=&\frac{K^z}{u}-(u+\beta) I,\nn \\
\B(u)&=&\frac{K^-}{u}-c, \nn \\
\C(u)&=&c^\dagger-\frac{K^+}{u}, \nn \\
\D(u)&=&-\frac{K^z}{u},  \eea 
with 
$$\aa(u)=\frac{\kappa}{u}-u-\beta,~~~~~~\dd(u)=-\frac{\kappa}{u}. $$ 
This realisation is evidently not unitary, but it is clear that 
$$\l<\vec w,\kappa \r|=(-1)^M\l\{\vec w,\kappa\r| $$    
even for arbitrary $\{w_i\}$. 
Through using the quasi-classical limit of the Slavnov formula
(\ref{qcslavnov})  we may find the scalar product of the states in these
models 
\bea \l<\vec w,\kappa|\vec v,\kappa\r>
&=&(-1)^M\l\{\vec w,\kappa|\vec v,\kappa\r> \nn \\
&=& \frac{1}{\prod^M_{k>l}(v_k-v_l)
\prod^M_{i<j}(w_i-w_j)}\det \FF \nn \\
&=& \frac{\prod^M_{p=1}\prod^M_{q\neq
p}(v_p-w_q)}{\prod^M_{k>l}(v_k-v_l)
\prod^M_{i<j}(w_i-w_j)}\det \GG \label{amslavnov}      
\eea 
where we have defined   
\bea 
\FF_{ij}&=&-\F_{ij}\nn \\
&=&-\l(\aa(v_j)-\dd(v_j)
+\sum_{k\neq i}^M\frac{2}{v_j-w_k}\r)
\frac{\prod_{l=1}^M(v_j-w_l)}{(v_j-w_i)^2}\nn \\
\GG_{ij}&=&-\G_{ij}\nn \\
&=&-\l(\aa(v_j)-\dd(v_j)
+\sum_{k\neq i}^M\frac{2}{v_j-w_k}\r)\frac{(v_j-w_j)}{(v_j-w_i)^2} \nn 
\eea   
in order to absorb the factor
$(-1)^M$. The set $\{w_i\}$ provide a solution to the quasi-classical limit of
the Bethe ansatz equations 
\beq w_i+\beta-\frac{2\kappa}{w_i}=\sum^M_{k\neq i}\frac{2}{w_i-w_k} 
. \label{qcambae} \eeq 
In the quasi-classical limit the energy eigenvalues for the Hamiltonian
(\ref{acham}) are 
$$E(\vec v)= \mu_a(2M+2\kappa-1/2)-2\Omega\sum_{i=1}^M w_i,    $$ 
while for (\ref{abcham}) they are given by 
$$E(\vec v)=\mu_a(M+2\kappa-1)+\mu_b M-\Omega\sum_{i=1}^M w_i. $$ 
In deriving the above energy expressions we have used the identity 
$$\beta M+\sum_{i=1}^M w_i=\sum_{i=1}^M\frac{2\kappa}{w_i}$$ 
which follows from (\ref{qcambae}).  

For the case when $\{v_j\}$ also
satisfy the Bethe ansatz equations we find for the elements of $\FF$ 
\bea \FF_{ij}&=&
-\l(\frac{2\kappa}{v_j}-v_j-\beta 
+\sum^M_{k\neq i} \frac{2}{v_j-w_k}\r)
\frac{\prod_{l=1}^M(v_j-w_l)}{(v_j-w_i)^2} \nn \\
&=&
-\l(2\kappa\l(\frac{1}{v_j}-\frac{1}{w_i} 
\r)+w_i-v_j+\sum_{k\neq i}^M\frac{2}{v_j-w_k}-\sum^M_{k\neq
i}\frac{2}{w_i-w_k} \r) \nn \\
&&~~~~~~~~~\times \frac{\prod^M_{l=1}(v_j-w_l)}{(v_j-w_i)^2} \nn \\
&=&\l(1+\frac{2\kappa}{w_iv_j}
+\sum^M_{k\neq i}\frac{2}{(w_i-w_k)(v_j-w_k)}\r)
\prod^M_{l\neq i}(v_j-w_l)\label{amf} \eea   
and similarly 
\beq \GG_{ij}=\l(1+\frac{2\kappa}{w_iv_j}
+\sum^M_{k\neq i}\frac{2}{(w_i-w_k)(v_j-w_k)}\r)
\frac{(v_j-w_j)}{(v_j-w_i)}. \label{amg} \eeq  
Letting $\l|\vec v,\kappa\r>= \l|\vec w,\kappa\r>$ gives us the square of
the norm formula 
$$ ||\vec v||^2=\det \KKK $$   
where 
\bea \KKK_{ii}&=&1+\frac{2\kappa}{v_i^2}+\sum_{k\neq i}^M 
\frac{2}{(v_i-v_k)^2} \nn \\
\KKK_{ij}&=&-\frac{2}{(v_i-v_j)^2} ~~~~~~~~~
~~~~~~~~~~~~~~{\rm for}~~~~~i\neq j. \nn \eea 

To compute the form factors for $K^z$, we need to take a limit of  
the form factors of $\D(u)$ as given by (\ref{qcmed}). 
This leads us to 
\bea \l<\vec w,\kappa|K^z|\vec v,\kappa\r>&=&-\lim_{u\rightarrow 0}
u\l<\vec w,\kappa|\D(u)|\vec v,\kappa\r> \nn \\
&=&\frac{\kappa\prod_{k=1}^Mw_k}
{\prod_{k>l}^M(v_k-v_l)\prod_{i<j}^M(w_i-w_j)\prod_{k=1}^M v_k}\det(\FF-\QQ)
\nn \eea
with
\bea
\QQ_{ij}&=&-\frac{2\prod^M_{l\neq j}(v_j-v_l)}{w_i^2}. \nn \eea
Using the fact that $K^z+N_c$ is conserved in both models
(\ref{acham},\ref{abcham}) and can be expressed in terms of the total atom 
number $N$, the form factors for $N_c$ can be deduced from those
for $K^z$. 

Next we turn to the problem of finding the form factors for the
operators $c^\dagger,\,c,\,K^+$ and $K^-$. To do this
we need to solve the inverse problem and express each of these operators
in terms of the realisation of the Gaudin algebra. This is not difficult
to achieve with the result 
\bea 
K^+&=&-\lim_{u\rightarrow 0}u\C(u),  \nn \\
K^-&=&\lim_{u\rightarrow 0}u\B(u), \nn \\
c^\dagger&=& \lim_{u\rightarrow \infty}\C(u), \nn \\
c&=&-\lim_{u\rightarrow \infty}\B(u) . \label{aminverse} \eea 

Using the fact that the parameters $\{v_i\}$ in the Slavnov formula 
(\ref{amslavnov}) are arbitrary, we can then take the limits described
above to yield the form factors. This gives the results
\bea 
\l<\vec v,\kappa|K^-|\vec w,\kappa\r>
&=&\l<\vec w,\kappa|K^+|\vec v,\kappa\r> \nn \\
&=&-\lim_{u\rightarrow 0}u\l<\vec w,\kappa|v_1,\cdots,v_{M-1},u,\kappa\r> \nn
\\
&=& \frac{\prod_{q=1}^Mw_q}
{\prod_{k>l}^{M-1}(v_k-v_l)
\prod_{i<j}^M(w_i-w_j)\prod_{p=1}^{M-1}v_p}\det \P\nn \eea 
where 
\bea \P_{ij}&=&\FF_{ij} ~~~~~~~{\rm for}~~~j\neq M, \nn \\
\P_{iM}&=&-\frac{2\kappa}{w_i^2},               \nn \eea 
and
\bea
\l<\vec v,\kappa|c|\vec w,\kappa\r>
&=&\l<\vec w,\kappa|c^\dagger|\vec v,\kappa\r> \nn \\
&=&\lim_{u\rightarrow \infty}\l<\vec w,\kappa|v_1,\cdots,v_{M-1},u,\kappa\r> 
\nn \\
&=& \frac{\prod_{p=1}^{M-1}\prod_{q\neq p}^M(v_p-w_q)}
{\prod_{k>l}^{M-1}(v_k-v_l)
\prod_{i<j}^M(w_i-w_j)}\det \W\nn \eea
where
\bea \W_{ij}&=&\GG_{ij} ~~~~~~~{\rm for}~~~j\neq M, \nn \\
\W_{iM}&=&1.              \nn \eea

\subsection{Form factors for the reduced BCS model} 

The results of this section have been published in \cite{zlmg} for the
case $s_k=1/2,\,\forall\,k$ (although different conventions and
notations were used). A
closely related study is given in \cite{ao}. The fundamental difference
between \cite{ao} and the results below is that \cite{ao} employs the
generating function of correlators of the Gaudin algebra as developed
in \cite{sklyanin2}, whereas below we will directly 
use the quasi-classical limit
of the Slavnov formula as given by (\ref{qcslavnov}). 
By this procedure
the form factors are obtained in an explicit determinant
representation. Again, we will derive results for the general case of the
Hamiltonian (\ref{ham1}) where the irreducible  
realisations of the $su(2)$ algebras, labelled by a lowest weight
$-s_k$, are arbitrary.
The realisation of the Gaudin algebra obtained by taking the
quasi-classical limit of the realisation of the 
Yang-Baxter algebra given by (\ref{bcstm}) reads
\bea
\A(u)&=&-\alpha I+\sum_{k=1}^\L\frac{S_k^z}{u-\e_k},\nn \\
\B(u)&=&\sum_{k=1}^\L\frac{S_k^-}{u-\e_k}, \nn \\
\C(u)&=&\sum_{k=1}^\L\frac{S_k^+}{u-\e_k}, \nn \\
\D(u)&=&\alpha I-\sum_{k=1}^\L\frac{S_k^z}{u-\e_k},  \eea
with
$$\aa(u)=-\alpha-\sum_{k=1}^\L\frac{s_k}{u-\e_k},~~~~~~
\dd(u)=\alpha+\sum_{k=1}^\L \frac{s_k}{u-\e_k}. $$
This realisation is unitary, so we do not need to deal with the issues
of non-unitarity as in the previous examples. 

Through using the quasi-classical limit of the Slavnov formula
(\ref{qcslavnov})  we find the scalar product of the states 
\bea \l<\vec w|\vec v\r>
&=& \frac{1}{\prod^M_{k>l}(v_k-v_l)
\prod^M_{i<j}(w_i-w_j)}\det \F\nn \\
&=& \frac{\prod^M_{p=1}\prod^M_{q\neq
p}(v_p-w_q)}{\prod^M_{k>l}(v_k-v_l)
\prod^M_{i<j}(w_i-w_j)}\det \G \label{bcsslavnov} \eea
where $\G_{ij}$ and $\F_{ij}$ are given by 
(\ref{qcg},\ref{qcf}) 
and the Bethe ansatz equations for the parameters $\{w_i\}$ are 
given by (\ref{bcsbae}). 
Letting $\{v_i\}$ also be a solution of the Bethe ansatz equations 
we find 
\bea \F_{ij}&=&
\l(-2\alpha-\sum_{k=1}^\L\frac{2s_k}{v_j-\e_k}
+\sum^M_{k\neq i} \frac{2}{v_j-w_k}\r)
\frac{\prod_{l=1}^M(v_j-w_l)}{(v_j-w_i)^2}
\nn \\
&=&\l(\sum_{k=1}^\L\frac{2s_k}{(w_i-\e_k)
(v_j-\e_k)}   
-\sum^M_{k\neq i}\frac{2}{(w_i-w_k)(v_j-w_k)}\r)
\prod_{l\neq i}^M(v_j-w_l),\nn \\
\G_{ij}&=&\l(\sum_{k=1}^\L\frac{2s_k}{(w_i-\e_k)
(v_j-\e_k)}
-\sum^M_{k\neq i}\frac{2}{(w_i-w_k)(v_j-w_k)}\r)
\frac{(v_j-w_j)}{(v_j-w_i)}. \nn \eea

Setting $\l|\vec v\r>= \l|\vec w\r>$ gives us the square of
the norm formula
$$ ||\vec v||^2=\det \KK $$
where
\bea \KK_{ii}&=&\sum_{k=1}^\L\frac{2s_k}{(v_i-\e_k)^2}-\sum_{k\neq i}^M
\frac{2}{(v_i-v_k)^2} \nn \\
\KK_{ij}&=&\frac{2}{(v_i-v_j)^2} ~~~~~~~~~
~~~~~~~~~~~~~~{\rm for}~~~~~i\neq j. \nn \eea
Putting $s_k=1/2,\,\forall\,k$, the above is exactly the norm square formula
obtained by Richardson \cite{richardson2}. 

To compute the form factors for $S^z_m$, we use the fact that 
$\D(u)$ has simple poles at $u=\e_j,\,\forall\, j$; i.e.
$$S^z_m= -\lim_{u\rightarrow \e_m}(u-\e_m)\D(u).$$
This leads to 
\bea \l<\vec w|S_m^z|\vec v\r>&=&-\lim_{u\rightarrow \e_m}
(u-\e_m)\l<\vec w|\D(u)|\vec v\r> \nn \\
&=&\frac{-s_m\prod_{k=1}^M(w_k-\e_m)}{\prod_{k=1}^M
(v_k-\e_m)
\prod_{k>l}^M(v_k-v_l)\prod_{i<j}^M(w_i-w_j)}\det(\F-\Q(\e_m))\nn  \eea
where $\F$  is as above and  
$$\Q_{ij}(u)=\frac{2\prod^M_{l\neq j}(v_j-v_l)}{(u-w_i)^2}. $$ 

Now we derive the form factors for the
operators $S^+$ and $S^-$. In this instance the inverse problem is
solved as follows 
\bea S^-_m&=& \lim_{u\rightarrow \e_m}(u-\e_m)\B(u), \nn \\
S^+_m&=&\lim_{u\rightarrow \e_m}(u-\e_m)\C(u). 
\label{bcsinverse} \eea
Using the fact that the parameters $\{v_i\}$ in the Slavnov formula
(\ref{bcsslavnov}) are arbitrary, we can then take the limits described
above to yield the form factors. The results are 
\bea
\l<\vec v|S_m^-|\vec w\r>&=&\l<\vec w|S_m^+|\vec v\r> \nn \\
&=&\lim_{u\rightarrow \e_m}(u-\e_m)\l<\vec w|v_1,\cdots,v_{M-1},u\r> \nn
\\
&=& \frac{\prod_{q=1}^M(w_q-\e_m)}
{\prod_{p=1}^{M-1}(v_p-\e_m)\prod_{k>l}^{M-1}(v_k-v_l)
\prod_{i<j}^M(w_i-w_j)}\det \P\nn \eea
where
\bea \P_{ij}&=&\F_{ij} ~~~~~~~~~~~~{\rm for}~~~j\neq M, \nn \\
\P_{iM}&=&\frac{1}{(w_i-\e_m)^2}.                \nn \eea

The above form factors can be used to construct general correlation
functions, such as the Penrose-Onsager-Yang off-diagonal long-range
order parameter as given in \cite{zlmg}. 

\section{Conclusion} 

We have reviewed the theory of the quantum inverse scattering method and
algebraic Bethe ansatz for the computation of energy spectra and form
factors in exactly solvable models, and demonstrated how it 
applies to several models of Bose-Einstein condensates and the reduced
BCS model.
Throughout we have only used the specific example of the Yang-Baxter
algebra associated with the Lie algebra $gl(2)$. However, a 
Yang-Baxter algebra can be associated with any simple Lie
algebra, Lie superalgebra, and the $q$-deformations of these structures.
Hence the theory can be applied on a much wider level. For example,
generalised BCS systems derived from Yang-Baxter algebras associated with the 
Lie algebras $gl(4)$ and $so(5)$ and the Lie superalgebra $gl(2|1)$ 
were derived in \cite{gflz,lzgm,hl} 
respectively. The model obtained in \cite{lzgm} was proposed by
Richardson in 1966 to describe proton-neutron pairing in nuclear systems
\cite{r66,r66a}. The case of a general Lie algebra
was examined in \cite{afs} in the context of the Knizhnik-Zamolodchikov
equation. 
One major challenge which remains is to extend the Slavnov formula for
the scalar products of states to the general case. In fact there has
been little progress on this aspect with the exceptions of the work by
Reshetikhin on the norms of the wavefunctions for models derived from
the  
$gl(3)$ Yang-Baxter algebra \cite{r} and 
by G\"ohmann and Korepin on the Hubbard
model \cite{gk}. Another approach based on the Knizhnik-Zamolodchikov
equation for $gl(n)$ can be found in \cite{tv}. 

In the method we have described the exact solution is parameterised in
terms of the Bethe ansatz equations, which cannot be solved analytically.
Consequently numerical analysis of the solutions must be undertaken. For
the BCS model there have been quite a number of works on this topic
(e.g. see \cite{vdr,admor,ao,sddvb,rsd}). 
For the models of Bose-Einstein condensates the only numerical
analysis of an exact solution,  of which we know, is in 
\cite{zlm}. It is also possible to
conduct an asymptotic analysis of the Bethe ansatz equations to find the
exact asymptotic behaviour of the energy spectrum and correlation
functions. Examples are given in \cite{flz,zlmx,zlm}.  

\begin{flushleft}
Acknowledgements-
Financial support from the Australian Research Council is gratefully
acknowledged. We thank Angela Foerster and Vladimir Korepin for correspondence.
\end{flushleft}

\section*{References} 


\end{document}